\documentclass[preprint,12pt]{elsarticle}

\usepackage[dvipsnames]{xcolor}
\usepackage[T1]{fontenc}
\usepackage[utf8]{inputenc}
\usepackage{xcolor}
\usepackage{amsfonts}
\usepackage{mathrsfs}
\usepackage{amsmath}
\usepackage{pifont}
\usepackage{booktabs}  
\usepackage{caption}
\usepackage{subcaption}
\usepackage{multirow}  
\usepackage{float} 
\usepackage{dsfont}
\usepackage{hyphenat}
\usepackage{cancel}
\usepackage{svg}
\usepackage{lscape}
\usepackage{graphicx}
\usepackage{longtable}
\usepackage{rotating}
\usepackage{booktabs}
\usepackage{makecell}
\usepackage{geometry}
\usepackage{orcidlink}

\usepackage[numbers]{natbib} 

\usepackage{adjustbox}
\usepackage{fancybox,framed}
\usepackage{tcolorbox}

\usepackage{hyperref}
\usepackage{comment}
\hypersetup{
    colorlinks=true,
    linkcolor=blue,
    filecolor=magenta,      
    urlcolor=blue,
    pdftitle={Overleaf Example},
    pdfpagemode=FullScreen
    }

\begin{document}
\begin{frontmatter}
\title{An MRI-informed poromechanical model for organ-scale prediction of glioma growth}

\author[1]{M. Abbad Andaloussi\orcidlink{0009-0003-9274-2086}}
\author[2]{S. Urcun\orcidlink{0000-0002-5164-5904}}
\author[3,4]{D. A. Hormuth II}
\author[5,3]{G. Lorenzo}
\author[6,7,8]{G. Sciumè}
\author[3,9,10,11,12]{C. Wu}
\author[3,4,9,13,14]{T. E. Yankeelov}
\author[1]{S. P. A. Bordas}

\address[1]{Faculty of Science, Technology and Medicine, University of Luxembourg,  University of Luxembourg, Belvaux, Luxembourg}
\address[2]{Inria Research Center at Rennes University, Campus de Beaulieu, 263 Av. Général Leclerc, Rennes, 35042, France}
\address[3]{Oden Institute for Computational Engineering and Sciences, The University of Texas at Austin, Austin, TX, USA}
\address[4]{UT Austin Cancer Research Institute, The University of Texas at Austin, Austin, TX, USA}
\address[5]{Group of Numerical Methods in Engineering, Department of Mathematics, and CITEEC, University of A Coruña, Spain}
\address[6]{Université de Bordeaux, CNRS, Bordeaux INP, I2M, UMR 529 , France}
\address[7]{Arts et Metiers Institute of Technology, CNRS, Bordeaux INP, Hesam Universite, I2M, UMR 5295, France}
\address[8]{Institut universitaire de France (IUF), France}
\address[9]{Department of Imaging Physics, The University of Texas MD Anderson Cancer Center, Houston, TX, USA}
\address[10]{Department of Breast Imaging, The University of Texas MD Anderson Cancer Center, Houston, TX,USA}
\address[11]{Department of Biostatistics, The University of Texas MD Anderson Cancer Center, Houston, TX, USA}
\address[12]{Institute for Data Science in Oncology, The University of Texas MD Anderson Cancer Center, Houston, TX, USA}
\address[13]{Department of Biomedical Engineering, The University of Texas at Austin, Austin, TX, USA}
\address[14]{Department of Diagnostic Medicine, The University of Texas at Austin, Austin, TX, USA}

\begin{abstract}

Gliomas constitute one of the most aggressive and heterogeneous forms of brain tumors, posing major challenges for understanding their biology and developing effective treatments.  Animal models enable the collection of rich longitudinal datasets describing tumor dynamics, which can be integrated within mathematical models to elucidate the biological mechanisms governing tumor growth. While most formulations rely on reaction-diffusion systems with limited insight on tissue deformation and fluid transport, we propose a magnetic resonance imaging (MRI)-informed, poroelastic model to describe C6 glioma growth in rats. We use data from animals ($n=4$) that were imaged five times after intracranial injection of cancer cells. Each MRI dataset includes (i) anatomical $T_1$-weighted data for brain and tumor segmentation and to assign mechanical properties; (ii) diffusion-weighted MRI, which enables estimation of the fraction of each voxel that is tumor; and (iii) dynamic contrast-enhanced MRI, which informs permeability as well as vascular and liquid fraction maps. Using finite-element simulations, model calibration for each rat uses the Levenberg-Marquardt method informed by the first three MRI datasets. Tumor forecasts are validated by assessing model-data agreement on the remaining two MRI datasets. Our results show relative tumor volume errors between 0.94 \% and 11.27 \% at calibration, and prediction errors between 4.73 \% and 36.03 \%.  Additionally, Dice scores ranged from 0.80 to 0.93 during calibration, and from 0.75 to 0.93 during validation. Thus, our results suggest that our poromechanical model can describe C6 glioma growth. This study provides a first step toward a patient‑specific, multiscale model of the spatiotemporal poromechanics underlying glioma progression and therapeutic response.

\end{abstract}

\begin{keyword}
Magnetic resonance imaging \sep finite elements \sep poromechanics \sep cancer \sep glioma \sep computational oncology

\end{keyword}

\end{frontmatter}

\section{Introduction}

Gliomas are a diverse group of primary brain tumors that arise from glial cells, which provide structural and metabolic support to neurons in the central nervous system \cite{waker2019brain, castillo2010stem,weller2015glioma, jiang2012origin, holland2001progenitor}. Although high-grade gliomas (e.g., glioblastomas) are inherently malignant, low-grade gliomas may also progress over time and exhibit a more aggressive behavior through a malignant transformation \cite{frazier2010rapid,satar2021systematic}. Due to their pronounced heterogeneity and rapid dynamics, the clinical study of gliomas is particularly challenging using the limited data that can be safely collected from human patients during the course of the clinical management of their disease \cite{comba2021uncovering}. Alternatively, animal models of glioma are experimental systems in which tumors are induced or implanted in animals (e.g., rats) to replicate key features of the disease in vivo \cite{grobben2002rat,lenting2017glioma,hicks2021large}. Hence, animal models allow for the controlled investigation of tumor behavior and treatment response. 
Accurate prediction of glioma growth dynamics is essential for effective treatment planning and intervention. This capability could serve to guide personalized surgical and radiation-based interventions \cite{lipkova2019personalized,corwin2013toward, hormuth2025forecasting}, as well as to plan fruitful experimental plans leading to discovery of novel biological mechanisms underlying glioma growth and treatment response \cite{gevertz2024minimally, segura2022optimal}. Towards these ends, several mathematical models have been developed to describe the dynamics of glioma growth and treatment response including tumor cell proliferation, invasion of neighboring healthy tissue, and response to therapies \cite{mang2020integrated,urcun2022oncology}. Nevertheless, patient-specific forecasting of the spatiotemporal dynamics of tumor growth in gliomas remains a signiﬁcant challenge due to the limited imaging and clinical data collected for each patient in the standard of care \cite{stupp2007chemoradiotherapy,weller2013standards}. To overcome this limitation, longitudinal datasets collected from experimental studies with animal models offer an information-rich context in which mathematical models of glioma growth can be constructed, tested for tumor forecasting, and further developed to operate in data-scarce conditions in the clinic (e.g., constructing simpler formulations that can be parameterized with less data without compromising predictive accuracy) \cite{hormuth2015predicting,hormuth2017mechanically,rey2024heterogeneous, benzekry2014classical}.

The majority of models to describe experimental or clinical glioma growth and their response to treatments rely on either temporally-resolved formulations based on ordinary differential equations (ODEs) \cite{plaszczynski2023predicting,bruningk2021intermittent,delobel2023overcoming} or spatiotemporal reaction-diffusion systems of partial differential equations (PDEs) based on the Fisher-Kolmogorov model \cite{wang2009prognostic, jackson2015patient, hormuth2017mechanically, balcerak2025individualizing, lipkova2019personalized, harkos2022inducing}. The literature on mathematical modeling of gliomas further includes phase-field modeling approaches \cite{agosti2018personalized, lima2017selection}, which constitute an alternative spatiotemporal formulation for tumor growth different to the Fisher-Kolmogorov paradigm. Spatiotemporal PDE models of glioma have become an intense area of research due to the central role of magnetic resonance (MRI) in tumor diagnosis, treatment planning, and patient monitoring \cite{rees2003advances,jenkinson2007advanced, sledzinska2024beyond,castellano2017functional}. Anatomical MRI sequences, such as $T_1$-weigthed and $T_2$-FLAIR (fluid Attenuated Inversion Recovery), enable segmentation of brain structures and delineation of the tumor region \cite{bhagat2018multiclass,hormuth2021image,bond2023image}. Quantitative sequences, like diffusion-weighted MRI (DW-MRI) and dynamic contrast-enhanced MRI (DCE-MRI), further allow for spatially probing biological features of healthy brain and tumor tissue, such as cellularity and vascularity, respectively \cite{lorenzo2022quantitative, hormuth2021image}.

Spatiotemporal PDE models of glioma have also accounted for the mechanical interaction between tumor cells and their microenvironment \cite{agosti2018personalized,harkos2022inducing,hormuth2017mechanically}, which have been shown to affect tumor dynamics and architecture \cite{chen2020multiscale,rejniak2010current,katira2013modeling,jain2014role}. However, there is a dearth of models that fully capture the biophysical coupling between tumor growth dynamics, tissue deformation, and fluid transport, which all together shape tumor behavior and treatment response \cite{harkos2022inducing,jain2014role}. In particular, poromechanical models are capable of explicitly modeling this complex interplay between tumor cells and surrounding brain tissue. Originally developed in soil mechanics \cite{selvadurai2021poroelastic, biot1941general, terzaghi1996soil}, these models can represent the multiphase nature of soft biological tissues, in which tumor cells, extracellular matrix (ECM), and interstitial liquid coexist \cite{sciume2014tumor, sciume2014three, sciume2021mechanistic}. Indeed, prior studies have shown that brain and tumor tissues can be approximated as poroelastic media leveraging various deﬁnitions of porosity based on the modeling assumptions. For example, while Urcun \emph{et al.} \cite{urcun2023non} modeled both tumor and healthy cells within the fluid phase, Rey \emph{et al.} \cite{rey2024heterogeneous} restricted porosity to the interstitial liquid only. However, despite the comprehensive biomechanical framework provided by poromechanics to represent glioma growth, most modeling efforts rely on a large number of parameters that cannot be reliably identified from clinical or experimental data. This is a crucial limitation that impedes their reliable application to aid in the discovery of biological mechanisms of glioma in experimental settings or guide patient-specific decisions through personalized tumor forecasting in clinical scenarios.

By exploiting the advantage of rich longitudinal data in experimental settings, here we propose an MRI-informed poromechanical model to describe untreated glioma growth in rats using a formulation that only depends on four driving parameters. We identify the latter through a sensitivity study, while maintaining a model formulation that accounts for intra- and intertumoral spatial heterogeneity in mechanical properties, vascularity, and tumor and liquid cell fractions. Spatial measurements from these model quantities were obtained from MRI scans of $n$=4 rats at five timepoints after the injection and initial growth of the glioma cells by using both anatomical and quantitative MRI methods. These data enabled animal-specific model initialization and calibration, as well as validation of tumor forecasts. Indeed, the preliminary tumor forecasting study conducted with our model in this work shows promising predictive capabilities, while also enabling a detailed assessment of critical biomechanical factors, including tissue deformation, interstitial fluid pressure, and stress distributions.

 The paper is structured as follows. First, we introduce the experimental data used to inform the model. The mathematical model is then presented, including a description of constitutive equations as well as their parameters and variables. We then describe the approach for sensitivity analysis to identify the model parameters driving glioma dynamics, as well as the method to optimally calibrate the parameters for each rat. We also describe the numerical approach, from spatial discretization to time integration. Then, we present our forecasting results, including tumor volume and solid pressure predictions. Finally, we discuss our results, study limitations, and future work.

\section{Materials and methods}

\subsection{Experimental data}
As the rats in this study represent a subset of those imaged in Ref. \cite{hormuth2015predicting}, here we only summarize the salient experimental details. We note that all experimental procedures were approved by the Institutional Animal Care and Use Committee at Vanderbilt University (Nashville, TN, USA). 
After intracranial injection of C6 tumor cells, each rat was imaged beginning 10 days post-surgery and up to 10 days after the first imaging time point. We begin our modeling study on the second MRI visit (day 12 post-surgery, henceforth labeled day 0) which had an average tumor volume greater than 109 mm$^3$, thereby enabling an accurate observation on MRI. The remaining imaging visits used were at days 2, 3, 4 and 6. 
\\
All MR images were acquired with a spatial resolution of 128 $\times$ 128 $\times$ 16 and then cropped to 61 $\times$ 41 $\times$ 16 to only contain the brain region. The dataset consisted of high-resolution $T_1$-weighted images, DCE-MRI, and DW-MRI. The high-resolution $T_1$-weighted, anatomical images were used to extract tumor and brain segmentations. The DCE-MRI was analyzed with the Tofts-Kety model \cite{barnes2012} to estimate the extravascular extracellular volume fraction map ($v_e$). Additionally, the relative cerebral blood volume (rCBV) was computed as the ratio of the area under the curve for the concentration of the contrast agent time course for each voxel to the area under the arterial input function (AIF) \cite{hormuth2014comparison} over the ﬁrst 60 s. The DW-MRI data was processed to obtain apparent diffusion coeﬃcient (ADC) maps. The ADC images were individually normalized with respect to the ventricular area to ensure comparable values at each measurement time point.

 \subsection{Mathematical model}

\begin{figure}
    \centering
    \includegraphics[width=1\columnwidth]{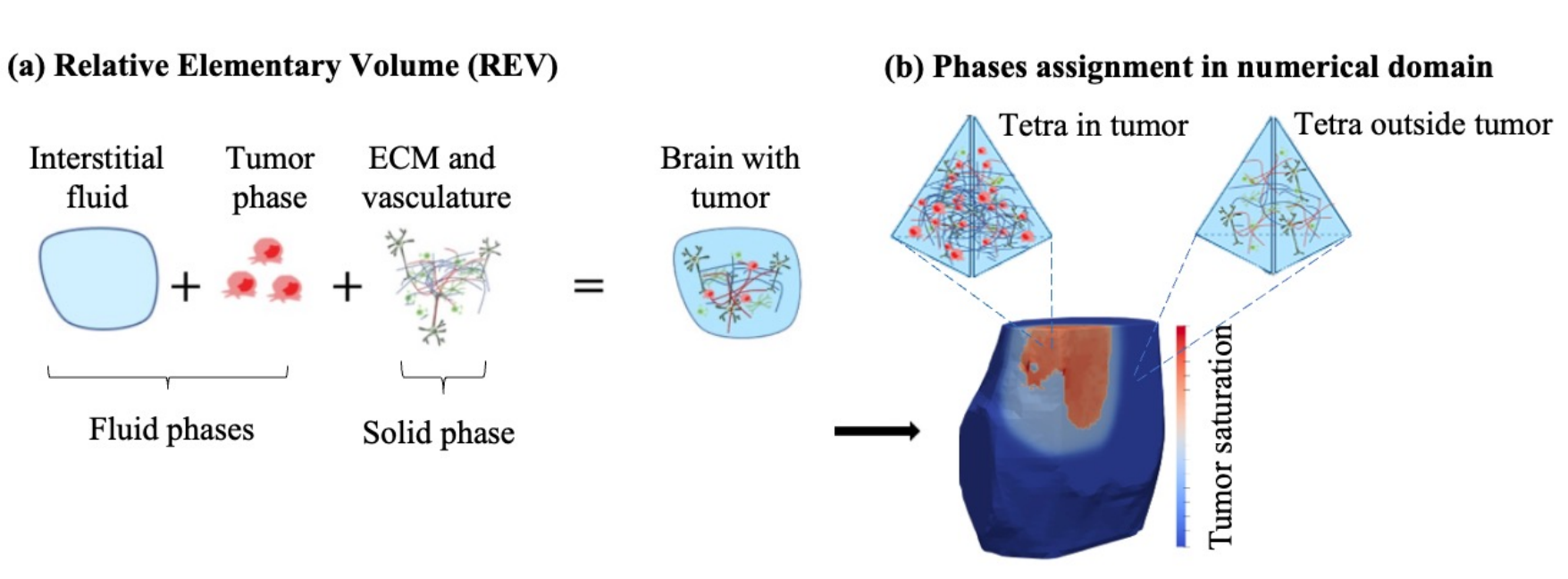}
    \caption{Definition of our three-phase model of glioma growth. (a) Components of one representative elementary volume (REV). Along with the interstitial liquid, the tumor phase is also considered a fluid phase that is part of the porosity. The extra-cellular matrix (ECM), brain healthy cells and the vascular system consitute the solid phase. (b) The numerical domain is defined from brain and tumor segmentations and tetrahedra's constituents are defined according to the REV. Variables such as tumor saturation are initialized from MRI into the numerical domain. The model parameters are subsequently calibrated and the model unknowns are predicted.
    }
    \label{REV}
\end{figure}

We propose a three-phase poroelastic imaging-informed model. 
The tissue is modeled as a three-phase continuum consisting of a solid scaffold, with volume fraction ($\varepsilon^{s}$), the interstitial liquid phase with volume fraction ($\varepsilon^{l}$), and the tumor cell phase with volume fraction ($\varepsilon^{t}$). We consider the latter two phases to be fluid phases within our poroelastic framework \cite{sciume2012tumor,urcun2021digital}. Therefore, the two fluid phases together with the solid scaffold are constituents of the representative elementary volume (REV) of the three-phase system (see Fig.~\ref{REV}). Additionally, the porosity $\varepsilon$ is calculated by summing the volume fraction of the fluid phases:

\begin{equation}
\varepsilon = \varepsilon^{l} + \varepsilon^{t}.
\label{poro1}
\end{equation} 

The solid phase, represented with the solid scaffold volume fraction ($\varepsilon^s$), consists of all remaining tissue constituents (i.e., ECM, neurons, vascular network, healthy glial and immune cells). The vascular network, providing oxygen and other nutrients, is modeled as a species (i.e., a sub-phase) of the solid phase represented with its mass fraction $\omega^{bs}$. 

Following from the definition of our three-phase system, the solid and fluid volume fractions must always satisfy the following constraint:

\begin{equation}
\varepsilon^{l} + \varepsilon^{t} + \varepsilon^{s} = 1.
\label{porosolid}
\end{equation}
From Eq.~\eqref{poro1}, we can also derive a saturation constraint as:
\begin{equation}
1 = S^{l} + S^{t},
\end{equation}
where each saturation $S^{f}=\frac{\varepsilon^f}{\varepsilon}$ $(f=l,t)$ represents the fraction of porosity that each fluid phase (i.e., interstitial liquid and tumor cells) occupies.

\subsubsection{Governing equations}\label{gov}

The mathematical model is governed by the conservation equations of mass and momentum of the considered phases.
To define these equations, we first introduce the material derivative of a physical quantity $f$  as:
 
\begin{equation}
    \dfrac{D^{s}(f)}{Dt} = \frac{\partial f}{\partial t} + \nabla{f}  \cdot {{\bf{v}}^{\overline s }}, 
    \label{4}
\end{equation}

\noindent which defines the temporal change of said physical quantity in a material element that is subjected to a spatiotemporal macroscopic velocity field. In our case, the deformable solid scaffold introduces the velocity field $\bf{v}^{\overline s}$ used in Eq.~\eqref{4}. Hence, our model will describe the motion of the fluid phases with respect to the deforming solid scaffold, which is considered as a reference frame.

\subparagraph{Mass conservation}
We assume that only the liquid and tumor phases exchange mass, such that the tumor takes its mass exclusively from the interstitial fluid to grow. The conservation of mass assumption leads to the following system of equations. 
The mass balance of the porous solid reads as:
\begin{equation}
    \dfrac{D^{s}(\rho^{s}\varepsilon^{s})}{Dt}+ \rho^{s}\varepsilon^{s} \nabla \cdot  {\bf{v}}^{\overline s} =0,
\end{equation}
with $\dfrac{D^{s}(\rho^{s}\varepsilon^{s})}{Dt}$ being the temporal evolution of the solid scaffold mass and $\rho^{s}\varepsilon^{s} \nabla \cdot  {\bf{v}}^{\overline s}$ being its deformation term. Assuming that the density of the solid scaffold $\rho^{s}$ is constant and replacing $\varepsilon^{s}$ according to Eq.~\eqref{porosolid}, we can write: 

\begin{equation}
    \dfrac{D^{s}\varepsilon}{Dt}=(1-\varepsilon )\nabla \cdot  {\bf{v}}^{\overline s}.
    \label{poro}
\end{equation}
\\ 
The mass conservation of the fluid phases can be written as:\\
\begin{equation}
    \dfrac{D^{s}(\rho^{t}\varepsilon S^{t})}{Dt} + \nabla  \cdot \left( {{\rho ^t}\varepsilon {S^t}{{\bf{v}}^{\overline {t s} }}} \right) + \;{\rho ^t}\varepsilon {S^t}\nabla  \cdot {{\bf{v}}^{\overline s }} \; =\qquad \mathop {\mathop M\limits_{} }\limits^{l \to t}, 
    \label{5}
\end{equation}

\begin{equation}
    \underbrace{\dfrac{D^{s}(\rho^{l}\varepsilon S^{l})}{Dt}}_{\text{Accumulation rate}} + \underbrace{\nabla  \cdot \left( {{\rho ^l}\varepsilon {S^l}{{\bf{v}}^{\overline {l s} }}} \right)}_{\text{flux term}} + \underbrace{{\rho ^l}\varepsilon {S^l}\nabla  \cdot {{\bf{v}}^{\overline s }}}_{\substack{\text{Solid deformation} \\ \text{contribution }}} = \underbrace{-\mathop {\mathop M\limits_{} }\limits^{l \to t},}_{\text{Mass transfer term}} 
    \label{6}
\end{equation}

where $\rho^{f}$ is the density of the fluid phase $f$, $\bf{v}^{ s }$ is the velocity of the solid phase, and ${\bf{v}}^{ {f s}}$ is the relative velocity of the fluid phase $f$ with respect to the solid phase $s$ defined as $\bf{v}^{ {f s}} = \bf{v}^{f}-\bf{v}^{s}$ with ${\bf{v}}^{ f }$ being the velocity of the fluid phase $f$ $(f={l,t})$. $\mathop {\mathop M\limits_{} }\limits^{l \to t}$ is the mass transfer term from phase $l$ to phase $t$.

\subparagraph{Momentum conservation}
The mass conservation equations are completed with the momentum balance of the system: 
\begin{equation}
   \nabla \cdot \mathbf{t}^{\Bar{\Bar{T}}}  = 0,
   \label{momentum}
\end{equation}
where $\mathbf{t}^{\Bar{\Bar{T}}}$ is the total Cauchy stress tensor. 
Each individual phase within the system is considered incompressible.
 As a result, the densities of the three phases are constant and the Biot coefficient, $\beta$, is 1. This allows us to write the effective stress principle as follows: 
\begin{equation}
  \mathbf{t}^{\Bar{\Bar{T}}}= \mathbf{t}^{\Bar{\Bar{E}}} - p^s \mathbf{\Bar{\Bar{1}}},
  \label{9}
\end{equation}
where $\mathbf{t}^{\Bar{\Bar{E}}}$ is the effective stress tensor, and $p^{s}$ is the solid pressure. The latter takes into account the contributions of the interstitial liquid and tumor weighted by their saturations:
\begin{equation}
  p^{s} = S^{l}p^{l} +S^{t}p^{t}.
  \label{10}
\end{equation}
~\\
\subparagraph{Constitutive relationships}

The three-phase porous continuum is considered linear elastic, with constant Poisson's ratio $\nu$ and a Young modulus taking different constant values in white and gray matter ($E_w$ and $E_g$, respectively). The effective stress tensor, $\mathbf{t}^{\Bar{\Bar{E}}}$, can therefore be defined according to Hooke's law such as:
\begin{equation}
    \mathbf{t}^{\Bar{\Bar{E}}} = \lambda tr(\epsilon) \mathbf{\Bar{\Bar{1}}} + 2 \mu \epsilon,
\end{equation}
where $\mathbf{\Bar{\Bar{1}}}$ is the identity tensor, $\epsilon(\mathbf{u^s}) = \frac{1}{2}(\nabla \mathbf{u^s}+ (\nabla\mathbf{u^s})^{T})$ is the linearized strain tensor while $\lambda = \frac{E \nu}{(1+\nu)(1-2\nu)}$ and $\mu = \frac{E}{2(1+ \nu)}$ are the Lamé's constants.
\\ \\
To describe the movement of the fluid phases through the porous medium, we use Darcy's law. Hence, we can define the relationship between the relative velocity, $\mathbf{v}^{f s}$ of the fluid phase $f$  ($f=l,t$) with respect to the solid phase $s$ and its corresponding pressure gradient, $\nabla p^{f}$, as:
    \begin{equation}
   - \frac{ k^s_{\text{int}}}{\mu^{f}}\nabla p^{f}  = \varepsilon S^{f}(\mathbf{v}^{f s}) \qquad {f}=t,l,
  \label{12}
    \end{equation}
where $k^s_{\text{int}}$ is the solid intrinsic permeability and $\mu^{f}$ is the dynamic viscosity of the phase $f$.
\\ \\ 
Each of the two fluid phases has its specific pressure $p^{f}$, so a pressure difference exists between the interstitial fluid and the tumor cell volume fractions. In our formulation, the interstitial fluid is considered as the wetting phase (i.e., the phase that preferentially sustains the load from the solid phase). Consequently, the pressure difference (the so-called capillary pressure) is defined as the difference between the pressure of the tumor cell phase and the interstitial fluid phase.

Following the assumptions of Sciumè \emph{et al.} \cite{sciume2014three}, and exploiting the parallelism with geophysics (\cite{parker1987model},\cite{parker1990determining}), the pressure jump between the two phases can be defined as: 
\begin{equation}
   \Delta p  = p^{t} - p^{l} = c \cdot\tan\left(\dfrac{\pi}{2}S^{t}\right).
  \label{12_bis}
    \end{equation}
Using Eqs.~\eqref{10} and \eqref{12_bis}, the solid pressure, $p^s$, can then be re-written as: 
\begin{equation}
 p^s=p^l + S^{t}c \cdot \tan\left(\dfrac{\pi}{2}S^{t}\right).  
 \label{ps}
\end{equation}
Additionally, the mass transfer term, $\mathop {\mathop M\limits_{} }\limits^{l \to t}$ has the following form:
\begin{equation}
    \mathop {\mathop M\limits_{} }\limits^{l \to t} = \mathop {\mathop \gamma\limits_{} }\limits^{l \to t} \varepsilon S^{t}(1-\varepsilon S^{t})(1-\mathcal{H}(p^{s},p_{\text{start}},p_{\text{crit}}))\omega^{bs},
    \label{masstransfer}
\end{equation}
where $\mathop {\mathop \gamma\limits_{} }\limits^{l \to t} $ is the tumor growth rate and $\varepsilon^s\omega^{bs}$ accounts for the supply of nutrients from the interstitial fluid when there is enough oxygen in the vasculature, which is informed using the \text{rCBV} (see section~\ref{fromMRI}). According to Eq.~\eqref{masstransfer}, tumor growth is controlled by three factors: the tumor phase, the interstitial liquid phase and the solid pressure. There is no mass transfer when there are no tumor cells (i.e., $S^{t} = 0$) or when the tumor saturates all local volume (i.e., $\varepsilon S^{t} = 1$).
Then, the function $\mathcal{H}(p^{s}, p_{\text{start}}, p_{\text{crit}})$ is a regularized heaviside function and accounts for how solid pressure controls the mechanics inhibition of tumor growth. This phenomenon starts once the solid pressure exceeds an initial threshold, $p_{\text{start}}$, and achieves its maximum intensity at a critical value, $p_{\text{crit}}$ when proliferation is fully inhibited. This relationship between tumor growth and solid pressure is embodied in the definition of the function $\mathcal{H}$:
    \begin{equation}
\mathcal{H}(p^{s}, p_{\text{start}}, p_{\text{crit}})=
 \left\{
   \begin{array}{l}
   \displaystyle 0 \quad\quad\quad\quad\quad\quad\quad\quad\quad\quad\quad\quad\text{if}\quad p^{s} \leq p_{\text{start}} \\
   ~\\
   \displaystyle  \frac{1}{2}-\frac{1}{2}\cos{\left(\pi\frac{p^{s}-p_{\text{start}}}{p_{\text{crit}}-p_{\text{start}}}\right)}\quad\text{if}\quad p_{\text{start}} \leq p^{s} \leq p_{\text{crit}}.\\
   ~\\
   \displaystyle 1 \quad\quad\quad\quad\quad\quad\quad\quad\quad\quad\quad\quad\text{if}\quad p^{s} \geq p_{\text{crit}} \\
   \end{array}
  \right.
  \label{chap5.29}
\end{equation}

\subsubsection{Primary model variables, boundary conditions, and initial conditions}
Based on the governing and constitutive equations introduced in Section~\ref{gov}, the model is controlled by three primary variables, namely: the tumor saturation ($S^t$), the solid phase displacement vector ($\mathbf{u^s}$), and the interstitial liquid phase pressure ($p^l$). Additionally, the porosity ($\varepsilon$) is used as a internal variable, while the rest of the model variables can be obtained by combining the model equations introduced in Section~\ref{gov}. This internal variable is calculated based on the update of the primary variables following equation~\eqref{4}.
Following previous models of brain cancer in the literature \cite{urcun2023non,mascheroni2018avascular,santagiuliana2018predicting}, we chose a non-homogeneous Dirichlet boundary condition for the interstitial liquid pressure ($p^l$=460 Pa), homogeneous Dirichlet boundary conditions for the solid displacements ($\mathbf{u^s}$=0), and a homogenenous Neumann boundary condition for the tumor saturation, meaning that there is no flux accross the boundary ($\nabla S^t \cdot \mathbf{n} = 0$).

We assume that there is no prior deformation (i.e., $\mathbf{u^s}(\mathbf{x},0)$=0), which is a common assumption in other poroelastic and mechanically-coupled models of tumor growth \cite{urcun2023non,blanco2023mechanotransduction,hormuth_abme}. The rationale for the initial conditions of the interstitial liquid pressure is as follows. Intracranial fluid pressure (IFP) has been extensively studied in various brain models of rats. Wiig \emph{et al.} measured IFP in a cohort of 83 rat brains using micropipettes and compared these values with cerebrospinal fluid (CSF) pressure. The average IFP was found to be approximately $457.31 \pm 86.66$ Pa, which is comparable to the CSF pressure of $471.41 \pm 94.71$ Pa \cite{wiig1983rat}. Following the IFP values in Wiig \emph{et al.}, we choose to use an initial IFP of $460$ Pa for our subsequent calculations.
While the interstitial liquid pressure and the solid displacement are initialized to the same values for all rats, the initial conditions for tumor saturation and porosity are defined from the MRI data collected at day 0 using the MRI-informed spatial maps for $\varepsilon^t$ and $\varepsilon^l$ (see Section~\ref{fromMRI} and Fig.~\ref{MRIinf}). 
Table~\ref{var_method} summarizes the definition of the primary variables in our model as well as their initialization and boundary conditions, along with supporting references.

\begin{table}[h!]
\centering
\begin{small}
\begin{tabular*}{\textwidth}{@{\extracolsep{\fill}}p{0.9cm} c c c c l@{}}
    \toprule
    \makecell{\textbf{Variable}} & \makecell{\textbf{Definition}} & \makecell{\textbf{Initial}\\ \textbf{conditions}} & \makecell{ \textbf{Boundary}\\ \textbf{conditions}} & \makecell{\textbf{Unit}} & \makecell{\textbf{References}} \\
    \midrule
    $S^{t}$& \makecell{Tumor \\ saturation} & \makecell{$\frac{\varepsilon^{t}}{\varepsilon}$\\in tumor region} & \makecell{$\mathbf{v}^{ts} \cdot \mathbf{n} = 0$} &[1] & \cite{majumder2023non, weissleder1998non, smits2014estimation} \\
        \midrule
    $\mathbf{u^s}$ & \makecell{Solid \\displacements} & $\mathbf{u^s}$=0 & 0 &\text{mm} & -  \\
        \midrule
    $p^{l}$ & \makecell{Interstitial \\ liquid pressure} & \makecell{460 } &\makecell{460} &\text{Pa} & \cite{wiig1983rat,boucher1997interstitial,elmghirbi2018toward} \\
    \bottomrule
\end{tabular*}
\caption{Definitions of the main model variables along with their boundary and initial conditions. The initial values of the tumor saturation field was computed from the tumor and interstitial liquid volume fraction fields (i.e., $\varepsilon^t$ and $\varepsilon^l$) measured from the first MRI dataset from each rat (see Section~\ref{fromMRI}).} \label{var_method}
\end{small}
\end{table}

\subsubsection{Constraining model variables and parameters with MRI measurements}\label{fromMRI}
\begin{figure}
  \centering

  \begin{subfigure}{0.19\textwidth}
    \includegraphics[width=\linewidth]{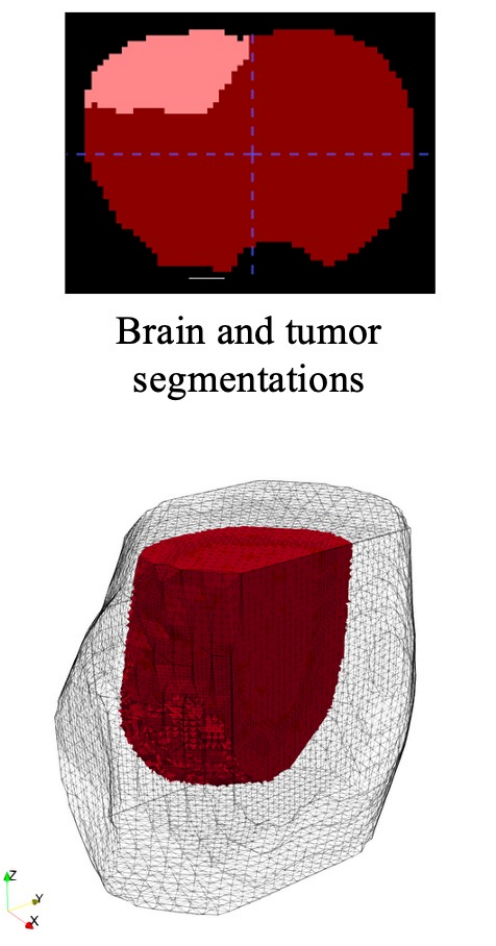}
    \caption{Brain and tumor 3D geometries}
    \label{fig:dispdiff:a}
  \end{subfigure}
  \hfill
  \begin{subfigure}{0.19\textwidth}
    \includegraphics[width=\linewidth]{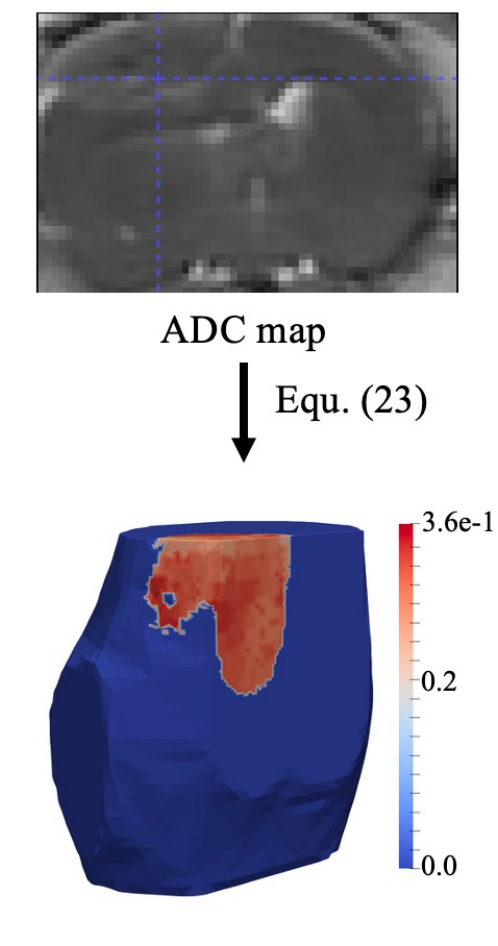}
    \caption{Tumor volume fraction}
    \label{fig:dispdiff:b}
  \end{subfigure}
  \hfill
  \begin{subfigure}{0.19\textwidth}
    \includegraphics[width=\linewidth]{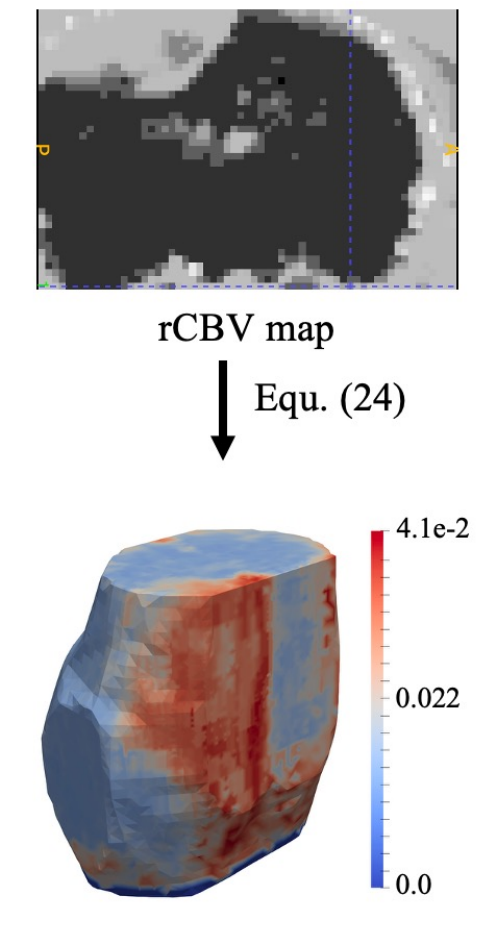}
    \caption{Vascular volume fraction}
    \label{fig:dispdiff:c}
  \end{subfigure}
  \hfill
  \begin{subfigure}{0.37\textwidth}
    \includegraphics[width=\linewidth]{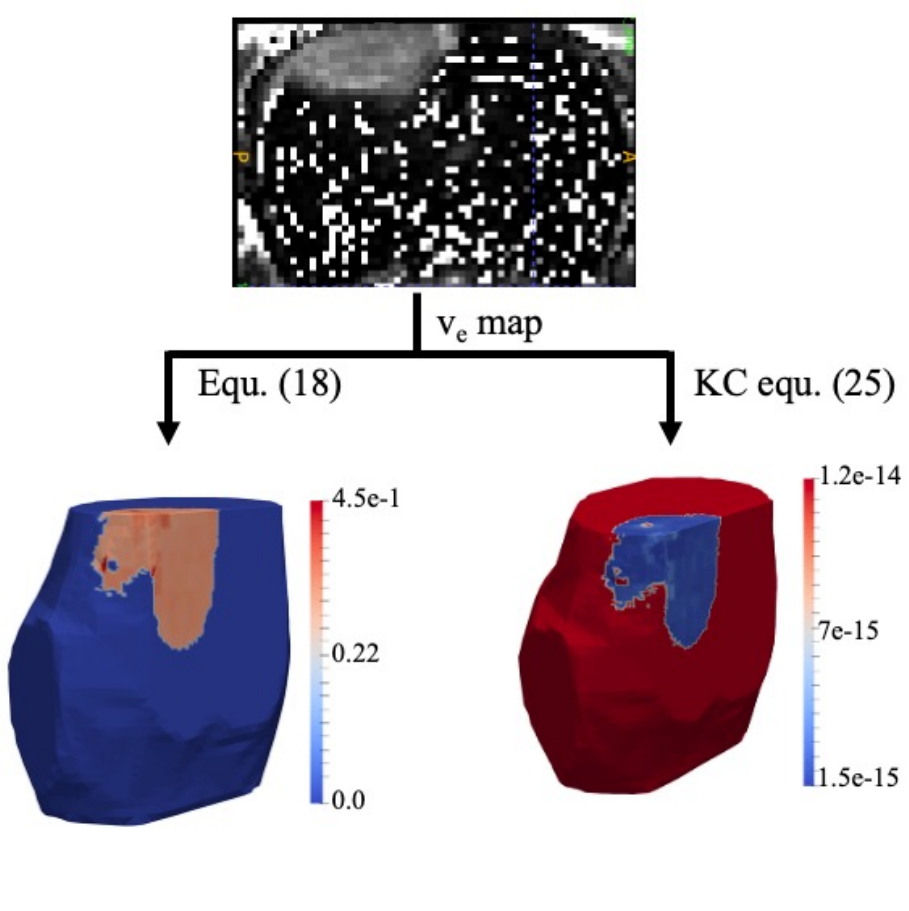}
    \caption{Interstitial liquid volume fraction (left) and permeability (right)}
    \label{fig:dispdiff:d}
  \end{subfigure}
    
  \caption{MRI-informed anatomical geometry and model variables. 
  (a) The computational 3D model of the brain and the 3D spatial region occupied by the tumor are defined from their corresponding segmentations on T1-weighted MRI (red and pink in the top row image, respectively). 
  (b) The tumor volume fraction ($\varepsilon^t$)  from the ADC values within the tumor region according to the linear inverse mapping in Eq.~\eqref{ADC}.
  (c) The vascular volume fraction ($\varepsilon^s\omega^{bs}$) was extracted from the rCBV maps within the whole domain by using Eq.~\eqref{wbs}. (d) The $v_e$ maps were used to inform both the interstitial fluid volume fraction by direct linear mapping \emph{via} Eq.~\eqref{equve} and permeability through Eq.~\eqref{KC} within the tumor region. Both the interstitial liquid volume fraction and the permeability were assigned uniform values outside the tumor (see Section~\ref{fromMRI}).}
  \label{MRIinf}
\end{figure}

The quantitative MRI measurements obtained from the rats enabled the definition of spatial maps for the model variables and some parameters within the brain geometry segmented on $T_1$-weighted MRI data. This process is described in detail in this section, and Table~\ref{fromMRI_table} summarizes the main imaging-based definitions of model components. Additionally, Fig.~\ref{MRIinf} illustrates how the MRI data is converted to model variables for the predictions of glioma dynamics with our model.

\begin{table}
\centering
\begin{small}
 \begin{tabular}{@{}p{1.6cm} c c c c l@{}}

    \toprule
    \textbf{Model component} & \textbf{Definition} & \textbf{Method} & \textbf{Value} & \textbf{Unit} & \textbf{References} \\
    \midrule
     $\varepsilon^{l}$ & \makecell{Interstitial \\ Liquid \\Volume \\ Fraction} & \makecell{Tumor: mapping \\from $v_e$ data \\ Healthy: literature}& \makecell{Tumor: Eq.~\eqref{equve}\\ \\ Healthy: 0.269} &\textbf{[1]} & \makecell{\cite{weissleder1998non,sykova2008diffusion} \\ \cite{lehmenkuhler1993extracellular,vendel2019need}}\\
         \midrule
    $\varepsilon^{t}$ & \makecell{Tumor \\Volume \\ Fraction} &  \makecell{Tumor: mapping \\from \textit{ADC} data \\ Healthy: literature} &\makecell{Tumor: Eq.~\ref{ADC}\\ \\Healthy: 0} &\textbf{[1]} & \cite{majumder2023non, weissleder1998non, smits2014estimation} \\
        \midrule
    $\varepsilon^s\omega^{bs}$ & \makecell{Blood \\ Supply \\Volume \\ Fraction} &\makecell{mapping \\from $rCBV$ data} & Eq.~\ref{wbs} &\textbf{[1]}  & \cite{pathak2001mr}\\
        \midrule
    $k^s_{int}$ & \makecell{Intrinsic \\ permeability} & \makecell{Tumor: mapping \\from $v_e$ data\\ Healthy: literature}& \makecell{Tumor: Eq.~\ref{KC}\\  \\Healthy: $1.22\times 10^{-14}$} & $\text{m}^{2}$ & \cite{truskey2004transport, majumder2023non, elmghirbi2018toward}\\
        \midrule
    $E^s$ & \makecell{Young's\\ Modulus} & \makecell{Values from literature, \\ spatial map from \\ $T_1$-weighted data} &\makecell{$E^s_{\text{grey matter}}$ = 344.9 \\ $E^s_{\text{white matter}}$ = 223.4} & $\text{Pa}$ &\cite{christ2010mechanical, finan2012viscoelastic, hormuth2017mechanically} \\
\bottomrule
  \end{tabular}
  \caption{ Definition of MRI-based model variables and parameters.}\label{fromMRI_table}
  \end{small}
  \end{table}

\subparagraph{Interstitial liquid volume fraction.} In the tumor tissue, we use the prior work by Weissleder \emph{et al.}, who obtained an interstitial volume fraction of $0.372 \pm 0.077$ through morphological measurements on C6-glioma slices \cite{weissleder1998non}. The minimum and maximum values are here linearly linked to $v_e$ map values following the equation:
\begin{equation}
    \varepsilon^l(x,y,z) =  (\varepsilon^l_{max}-\varepsilon^l_{min})v_e(x,y,z)+  \varepsilon^l_{min}.
    \label{equve}
\end{equation}

In healthy tissue, since the interstitial liquid is the only fluid phase, it equates to the porosity (i.e.,  $\varepsilon^l=\varepsilon$). In the literature, the porosity, also known as the extra-cellular, extra-vascular space, is approximately $20 \%$ in healthy rat brains \cite{lehmenkuhler1993extracellular,vendel2019need}. Sykova and Nicholson gathered further region-speciﬁc porosities in rat brains in a review \cite{sykova2008diffusion}. Based on these references, we estimated an average porosity value of $26.9 \%$ in healthy tissue.

\subparagraph{Tumor volume fraction} The tumor volume fraction map is defined by the following linear relationship based on the ADC map obtained from DW-MRI within the tumor segmentation \cite{hormuth2015predicting,urcun2023non, andaloussi2025malignant,fan2005usefulness}:

\begin{equation}
    \varepsilon^t(x,y,z) =  \frac{\varepsilon^t_{min}-\varepsilon^t_{max}}{ADC_{max}-ADC_{min}}ADC(x,y,z)+  \frac{\varepsilon^t_{min}ADC_{min}+\varepsilon^t_{max}ADC_{max}}{ADC_{max}-ADC_{min}},
    \label{ADC}
\end{equation}

\noindent where $ADC_{min}$ and $ADC_{max}$ are the minimum and maximum intensities of the ADC map for an individual rat and scan, respectively. Likewise, $\varepsilon^t_{min}$ and $\varepsilon^t_{max}$ are the minimum and maximum values of the tumor cell volume fraction in our framework, which we estimate from the literature as follows. Following the approach by Majumder et al. \cite{majumder2023non}, we can consider that the tumor tissue consists of cells $\varepsilon^c$ (which is the sum of healthy and tumoral cells, $\varepsilon^h$ and $\varepsilon^t$), interstitial space (structural elements of the extracellular matrix $\varepsilon^{ECM}$ and interstitial fluid, $\varepsilon^l$) and the vascular network $\varepsilon^b$. Bringing these elements together within our poroelastic framework, we obtain:
\begin{equation}
 \varepsilon^{ECM}+\varepsilon^{h}+\varepsilon^{b} +\varepsilon^{l}+\varepsilon^{t} = 1.
 \label{total}
\end{equation}
According to Majumder \emph{et al.} \cite{majumder2023non}, the average interstitial space volume fraction for different rat tumors is $0.503$. Thus, we can approximate $\varepsilon^{ECM}+\varepsilon^{l} = 0.503 $.  From the histological studies of Weissleder \emph{et al.} and Pathak \emph{et al.} \cite{weissleder1998non,pathak2001mr}, 
and by replacing $\varepsilon^{ECM}+\varepsilon^{l}$ as well as $\varepsilon^b$ by these approximate values from the literature, Eq.~\eqref{total} yields $\varepsilon^{t}+\varepsilon^{h} = 0.422 $ . 

We can now define minimum ( $\varepsilon^t_{min}$ ) and maximum ( $\varepsilon^t_{max}$ ) values of the tumor cell volume fraction to set bounds for the heterogeneous spatial map of $\varepsilon^t$ from MRI. Let us assume that $\varepsilon^t$ corresponds to the extreme case of a fully saturated tumor, such that there are no healthy cells (i.e., $\varepsilon^h =0$). By replacing $\varepsilon^h$ with 0, we estimate the maximum value for $\varepsilon^t$ as $0.422 $. On the other hand, the minimum percentage of tumor cells for histopathological tumor characterization is $\varepsilon^t_{min} = 0.001$ \cite{smits2014estimation}.

\subparagraph{Vascular volume fraction}  Pathak \emph{et al.} found a vascular volume fraction of $0.019 \pm 0.004$ on multiple histological cuts on 12 healthy rat brains \cite{pathak2001mr}. Additionally, the empirical comparison between maximum rCBV (healthy and tumor tissue intensities) and corresponding vascular volume fractions reported in the literature suggested a cubic-root relationship. Therefore, Eq.~\eqref{wbs} was adopted as a mapping between rCBV intensity and vascular volume fraction, where $rCBV_{max,h}$ normalizes the MRI signal to the maximum intensity in healthy tissue and $\varepsilon^s\omega^{bs,t}_{max}=0.023$ provides the physiological upper limit of the vascular volume fraction derived from histological measurements in the tumor tissue in Ref.~\cite{pathak2001mr}.
\begin{equation}
    \varepsilon^s\omega^{bs}(x,y,z)=\displaystyle \sqrt[3]{ \frac{rCBV(x,y,z)}{rCBV_{max,h}}}\varepsilon^s\omega^{bs,t}_{max}.
    \label{wbs}
\end{equation}


\subparagraph{Solid scaffold intrinsic permeability}
The Kozeny-Carman (\text{KC}) equation is widely used in the literature to link hydraulic conductivity and porosity. 
Hence, the solid intrinsic permeability can be deduced from 
\begin{equation}
    \frac{k^{s}_{int}}{\varepsilon} = \frac{\varepsilon^{3}}{C_{KC}S^2},
    \label{KC}
\end{equation}
where $C_{{KC}}$ is the \text{KC} constant, which we define as 5 as usually taken in literature for porosities smaller than $0.70$ \cite{truskey2004transport}. 
Additionally, $S=2/r$ is the wetted surface area per unit volume with $r$ being the radius of one cell.
In healthy tissue, the porosity only accounts for the interstitial liquid volume fraction, taking an average value of $\varepsilon = 0.269$, as explained above. 
Since the radius of a healthy cell is $r=5$~$\mu$m \cite{majumder2023non}, we estimate $k^s_{\text{int}}$ to be $1.22 \times 10^{-14}$ $\text{m}^2$ in healthy brain. 
In tumor tissue, Elmghirbi \emph{et al.} calculated a specific surface area corresponding to U251 tumor cells by incorporating the \text{KC} equation in Darcy's law and using a regression analysis \emph{versus} pressure, which resulted in $S=1443$ $\text{mm}^{-1}$  \cite{elmghirbi2018toward}.
Additionally, using Eqs.~\eqref{poro1}, \eqref{wbs}, and \eqref{ADC}, the spatial map of porosity ($\varepsilon$) in the tumor can be derived from the sum of the spatial maps of tumor volume fraction ($\varepsilon^t$) derived from $ADC$ data and the interstitial liquid volume fraction ($\varepsilon^l$) obtained from $v_e$ measurements, respectively.
As the ranges for $\varepsilon^t$ and $\varepsilon^l$ are $[0.001; 0.422]$ and $[0.295; 0.449]$ in tumor tissue, further using the aforementioned value of $S$ in tumor cells in Eq.~\eqref{KC}  we obtain a value range for the spatial map of intrinsic solid permeability in the tumor domain of $[1.46\times 10^{-15}, 1.28 \times 10^{-14}]$ $\text{m}^2$.

\subparagraph{Young's modulus} White- and gray-matter segmentations served to assign different Young's moduli values to these regions. An estimation of the Young's modulus was derived from previous experimental studies in the literature. For example, Finan \emph{et al.} \cite{finan2012viscoelastic} used microindentation techniques to determine the viscoelastic properties of various anatomical structures in sagittal slices of juvenile and adult rat brain. For adult rats, the shear modulus in the middle cortex was estimated to be approximately $1400 \text{Pa}$. Using a linear elastic law, the Young's modulus was subsequently estimated to be approximately $469 \text{Pa}$. 
Hormuth \emph{et al.} highlighted the spatial heterogeneity of shear modulus in the rat brain by assigning region-wise values: $G_{\text{cortex}} = 418 \text{Pa}$, $G_{\text{corpus callosum}} = 238 \text{Pa}$, $G_{\text{hippocampus}} = 466 \text{Pa}$,  $G_{\text{thalamus}} = 383 \text{Pa}$,  $G_{\text{putamen}} = 275 \text{Pa}$) \cite{hormuth2017mechanically}. Furthermore, Christ \emph{et al.} determined the apparent elastic modulus of rat cerebellum, distinguishing between white and gray matter. The elastic modulus for white matter was found to be $223,41 \pm 74 \text{Pa}$, and for gray matter $344.99 \pm 53 \text{Pa}$ \cite{christ2010mechanical}. These latter values were chosen for our computations.

\subsubsection{Homogeneous parameters}
Reference values for homogeneous parameters in the model were identified from the literature (see Table~\ref{param_method}).

\begin{table}
\centering
\begin{small}
 \begin{tabular}{@{}p{1.8cm} c c c c l@{}}

    \toprule
    \textbf{Parameter} & \textbf{Definition} & \textbf{Method} & \textbf{Value} & \textbf{Unit} & \textbf{References} \\
    \midrule
    $\nu$ & \makecell{Poisson's \\ ratio} & literature &\makecell{0.49} & - \\
        \midrule
    $\mu^l$ & \makecell{Interstitial \\ fluid \\ dynamic \\viscosity} & literature &\makecell{$3.5 \times 10^{-3}$} & $\text{Pa}\cdot \text{s}$ & \cite{yao2012interstitial,elmghirbi2018toward} \\
        \midrule
    $\mu^t$ & \makecell{Tumor cells \\ dynamic \\viscosity} & literature &[40-80] & $\text{Pa}\cdot \text{s}$ \\
        \midrule
    c & \makecell{Press-sat\\ parameter} & literature &[400-2000] & \text{Pa} & \cite{urcun2023non,sciume2013multiphase, sciume2014three} \\       \midrule
     $\mathop {\mathop \gamma\limits_{} }\limits^{l \to t}$ & \makecell{Tumor \\ growth rate} & literature &$5.29 \times 10^{-4}$ & $\text{kg}/\text{(m}^3\cdot\text{s)}$ & \cite{san1989assessment}\\
    \midrule
     $p_{\text{start}}$& \makecell{Start of\\ mechanical \\inhibition} & literature &1218 & \text{Pa}& \cite{urcun2023non}\\
    \midrule
     $p_{\text{crit}}$& \makecell{Critical \\mechanical \\inhibition} & literature &2680 & \text{Pa}&\cite{urcun2023non}\\
\bottomrule
  \end{tabular}
  \caption{ Definition of homogeneous model parameters.}\label{param_method}
  \end{small}
  \end{table}

\subparagraph{Tumor growth rate}
San-Galli \emph{et al.} assessed C6 glioma tumor size in mm$^{3}$ from 0 to 15 days in 60 Wistar rats \cite{san1989assessment}. From the size \emph{versus} time curve we estimate a growth rate of $5.29 \times 10^{-14}$ m$^3$/s.
Considering a tumor density of $\rho^{t}=1000$ kg$/$m$^{3}$, the growth rate in terms of mass is estimated to be $5.29 \times 10^{-11} kg/s$. 
Finally, since the average tumor volume in \cite{san1989assessment} is around $10^{-7} $m$^3$, we estimate a reference value of $\mathop {\mathop \gamma\limits_{} }\limits^{l \to t}$ as $5.29 \times 10^{-4}$ kg$/$(m$^3$.s).

\subparagraph{Interstitial liquid dynamic viscosity}
In the literature, $\mu^l$ ranges from $3.0 - 3.5 \times 10^{-3}$ $\text{Pa}\cdot\text{s}$ \cite{yao2012interstitial,elmghirbi2018toward}. We use the upper value of this range within this work.

\subparagraph{Tumor phase dynamic viscosity}
Following theoretical poromechanical formulations in the literature, the initial guess for tumor dynamic viscosity was set in the range $[40 - 100]$ Pa$\cdot$s \cite{sciume2014three, urcun2023non}.

\subparagraph{Pressure-saturation law parameter} Following previous poromechanical models using a similar law, the initial guess for the parameter c in Eq.~\eqref{ps} was chosen in the range $[500 - 1000]$.

\subparagraph{Poisson ratio} Following previous mechanical models characterizing brain tissue, the value $0.49$ was chosen for the Poisson ratio~\cite{hinrichsen2023inverse,urcun2023non,eskandari2021effect}.

\subparagraph{Mechanical inhibition pressure values $p_{start}$ and $p_{crit}$} Following previous poromechanical models for brain tumor models (glioma and glioblastoma), values of 1218 and 2680 were chosen for $p_{start}$ and $p_{crit}$ respectively \cite{urcun2023non,andaloussi2025malignant}.

\subsection{Numerical implementation}
\subsubsection{Mesh construction and sensitivity analysis}
The mesh was constructed using the tumor and brain segmentations on $T_1$-weighted MRI data. 
Both segmentations were exported as surface meshes in STL files to Meshlab for uniform mesh resampling and fix potential mesh issues (e.g., closing holes).
\cite{cignoni2008meshlab}. 
The brain STL file was subsequently used to construct a 3D mesh with tetrahedral elements by utilizing GMSH \cite{geuzaine2009gmsh}.
The tumor STL file was used to adaptively refine the rat brain mesh around the tumor volume with a raycasting algorithm from the Trimesh Python library \cite{trimesh} (see Fig.~\ref{MRIinf}).
The raycasting algorithm determines whether element midpoints lie inside the tumor by casting a ray from each midpoint parallel to the x-axis and checking if the number of intersections with the tumor surface is odd.
Subsequently, all tumor elements that share a face with a non-tumor element were marked as interface elements, representing the tumor border with surrounding healthy tissue.
The region of the mesh to be refined is defined by the volume of the tumor at the final time of a simulation using the initial parameter set for model calibration (see Section~\ref{Parameter_calibration}) on an unrefined mesh with homogeneous tetrahedron size.
Elements of the brain mesh in this region were assigned the first level of refinement (i.e., they are not refined with respect to the original brain mesh defined from imaging data).
Then, mesh refinements were performed gradually around the tumor boundary using the \textit{Dolfin} function refine \cite{logg2012dolfin} to define the second and subsequent levels of refinement.
To achieve an appropriate balance between accuracy and efficiency in our computations of glioma growth, we constructed six meshes with increasing local refinement around the external surface of the tumor (i.e., with two to six levels of refinement) for each rat, and we evaluated the predicted tumor volume at 2, 3 and 4 days. 
The tumor volume obtained with mesh 6 was considered the reference against which the corresponding result of all the other meshes was evaluated using the following error metric: 
\begin{equation}
    V_{i/6}=\frac{|V_i-V_6|}{V_6},
\end{equation}
with $i \in$ {1,2,3,4,5}.
\\

\subsubsection{Finite-element formulation}
We used the finite-element (FE) method to solve the coupled mass and momentum conservation equations governing the three-phase poromechanical model (see Section~\ref{gov}). 
Specifically, the momentum conservation (Eq.~\eqref{momentum}) solves for the solid displacement field $\mathbf{u^s}$, while the mass conservation equations solve for the tumor and interstitial liquid phases (Eqs.~\eqref{5} and \eqref{6}) and Eq.~\eqref{poro} updates the porosity at each time step.
We used a C++ code implementation of the FE method based on the Fenics legacy library \textit{Dolfin} \cite{alnaes2015fenics,logg2012dolfin}.
We first derived the weak formulation of the model equations introduced in Section~\ref{gov} (see ~\ref{Appendixvar} for details). 
First-order Lagrange elements were used for the scalar unknowns (i.e., the tumor saturation, the interstitial liquid pressure, and porosity) and second-order Lagrange elements were used for the vectorial unknown (i.e., the solid displacements).
The system was discretized in time using an implicit first-order backward Euler method. Nonlinearity in the governing equations was addressed \emph{via} the Newton-Raphson method using absolute and relative tolerances of $10^{-10}$ in the residual to assess convergence. To solve the linearized system emerging in each Newton-Rapshon iteration, we employed an LU direct solver.

The coupled system of mass and momentum conservation equations was first solved monolithically until the increment in the solid displacement field between two consecutive iterations fell below $10^{-4}$ mm. 
Subsequently, we used a staggered approach to reduce computation time. 
For the monolythic approach, the time step was progressively increased from 1 second during the first 30 iterations to 500 seconds by iteration 100. 
In the staggered scheme, the time step began at 0.1 seconds for the initial 10 iterations and was increased in four increments (1, 10, 60 and 250 seconds) until reaching 500 seconds by iteration 50. The algorithm iteratively alternated between solving the fluid subproblem (mass balance equations) and the solid subproblem (momentum balance equation) using the most recently computed variables from the other subsystem.
We observed that artifactual non-zero values of the tumor saturation may emerge in brain regions far from the tumor caused by numerical oscillations accumulating over successive time steps.
To avoid this issue, we reset any tumor saturation values below a threshold of $5\times10^{-6}$ to zero outside of the tumor.

\subsubsection{Variance-based parameter sensitivity analysis} 
A first-order variance-based sensitivity analysis was performed on a subset of the model parameters that were selected based on their relevance in controlling tumor dynamics in previous studies of computational oncology \cite{urcun2023non,urcun2021digital}, namely:  $\mu^t$, c, $\mathop {\mathop \gamma\limits_{} }\limits^{l \to t} $, $k^s_{int}$, $p_{start}$, and $p_{crit}$.
Since the intrinsic permeability ($k^{s}_{int}$) varies throughout the spatial domain, we introduce an auxiliary scalar coefficient $\alpha_k$ to uniformly scale this heterogeneous field and assess its global influence on model outputs. 
We followed the same methodology used in \cite{urcun2023non}, where more details on the method are introduced.
In brief, we estimated the Sobol indices to assess the sensitivity of the input parameters on the tumor volume, which is our main computational output. A cost function quantifying the error between the observed tumor volume (from tumor segmentations) and the numerical tumor volume (from domain regions where the model predicted $\varepsilon^{t} > 0.001$) was defined for each rat at day $3$. Each parameter was perturbed one at a time by $\pm50\%$ from its initial value and the corresponding volume variation was computed. The points of variation were linearly interpolated and the influence of a parameter is deduced from the slope of the linear fit. 
Finally, we defined the most sensitive parameters as those exhibiting larger first order Sobol indices and such that, together, gather $90 \%$ of the variance of the cost function.
These sensitive parameters were use to calibrate the model to each rat's longitudinal MRI data, while the rest of the parameters were fixed to constant values within their admissible range.

\subsection{Parameter calibration}
We leveraged the Gauss-Newton (GN) method to calibrate the sensitive model parameters for each rat using the MRI data up to day 3. Hence, we aimed at finding the optimal value of these parameters that minimize the error between the measured and model-estimated spatial extension of the tumor. The loss function of the GN method was computed as: 
\begin{equation}\label{gnloss}
    L(\varepsilon^t,\theta)=\frac{1}{2}\frac{\int( \mathds{1}\varepsilon^t_{num}-\mathds{1}\varepsilon^t_{obs})^2d\Omega}{\int(\mathds{1}\varepsilon^t_{obs})^2d\Omega},
\end{equation}
where $\theta$ is the set of parameters to optimize, $\varepsilon^t_{num}$ and $\varepsilon^t_{obs}$ are the model prediction and MRI measurement of the tumor volume fractions, and $\Omega$ is the computational brain domain.
In Eq.~\eqref{gnloss}, $\mathds{1} \varepsilon^t_{obs}$ and $\mathds{1} \varepsilon^t_{num}$
return $0$ outside the tumor region and $1$ inside for both the observation and numerical prediction. 
Following the histological observation that tumor is detected if there is at least $0.1 \%$ tumor cells in a tissue sample \cite{smits2014estimation}, the predicted tumor volume region is defined using the threshold $\varepsilon^t \geq 0.1 \%$.
The Jacobian matrix, which represents the sensitivities of model outputs to parameters within the GN method, was approximated by perturbing each parameter by $5 \%$.
To ensure that the parameters remained within physiologically plausible ranges during each GN iteration, upper and lower bounds were enforced as follows: $c \in [400, 2000]$, $\alpha_k \in [0.35, 4]$, $\mathop \gamma\limits_{} \limits^{l \to t} \in [5.29 \times 10^{-6}, 8.5\times 10^{-4}]$ $\text{kg}/\text{(m}^3\cdot\text{s)}$, $\mu^{t} \in [40, 100]$  $\text{Pa}\cdot\text{s}$. 
The stopping criteria for convergence consisted of either (i) achieving a reduction in the value of the parameter updates and in the residual error below a preset tolerance ($10^{-5}$, or (ii) reaching the upper or lower bound for the same parameter twice, meaning that there is no better parameter combination for the admissible range.
Additionally, we sampled 18 sets of initial guesses for the model parameters within the following ranges: $c \in [500, 1000]$, $\alpha_k \in [0.1, 1]$, $\mathop \gamma\limits_{} \limits^{l \to t} \in [5.29 \times 10^{-6}, 8.5\times 10^{-3}]$ $\text{kg}/\text{(m}^3\cdot\text{s)}$, $\mu^{t} \in [40, 90]$ $\text{Pa}\cdot\text{s}$ \cite{urcun2023non}. The set of initial guesses yielding the lowest error according to Eq.~\eqref{gnloss} was then selected as the starting point for the optimization procedure with the GN method.

\subsection{Quality of fits and forecasts}
We assess the quality of our model calibrations and predictions using the Dice score (DSC), the non-intersected volume (NIV) and the relative tumor volume error ($e_{vol}$), which are computed via:
\begin{equation}
    DSC = 2\frac{Vol_{num} \cap Vol_{obs} }{Vol_{num}+Vol_{obs}},
\end{equation}
\begin{equation}
    NIV(\varepsilon^t,\theta)=\frac{\int |\, (\mathds{1}\varepsilon^t_{num}-\mathds{1}\varepsilon^t_{obs})\,|d\Omega}{\int\mathds{1}\varepsilon^t_{obs}d\Omega},
\end{equation}
and 

\begin{equation}
    e_{vol} = \frac{|Vol_{num} - Vol_{obs}|}{Vol_{obs}},
\end{equation}

\noindent where $Vol_{num}$ and $Vol_{obs}$ are the tumor volumes from numerical predictions and MRI observations, respectively.

\section{Results}

\subsection{Mesh sensitivity analysis}
\begin{figure}[p]
  \centering
  \begin{subfigure}{0.8\textwidth}
    \includegraphics[width=\linewidth]{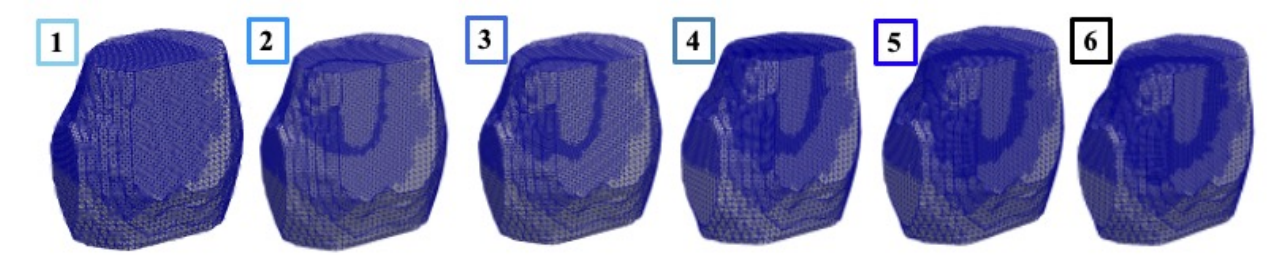}
    \caption{Meshes with different levels of refinement around the tumor for rat 1}
    \label{sensi1a}
  \end{subfigure}
  \hfill
  \begin{subfigure}{0.43\textwidth}
    \includegraphics[width=\linewidth]{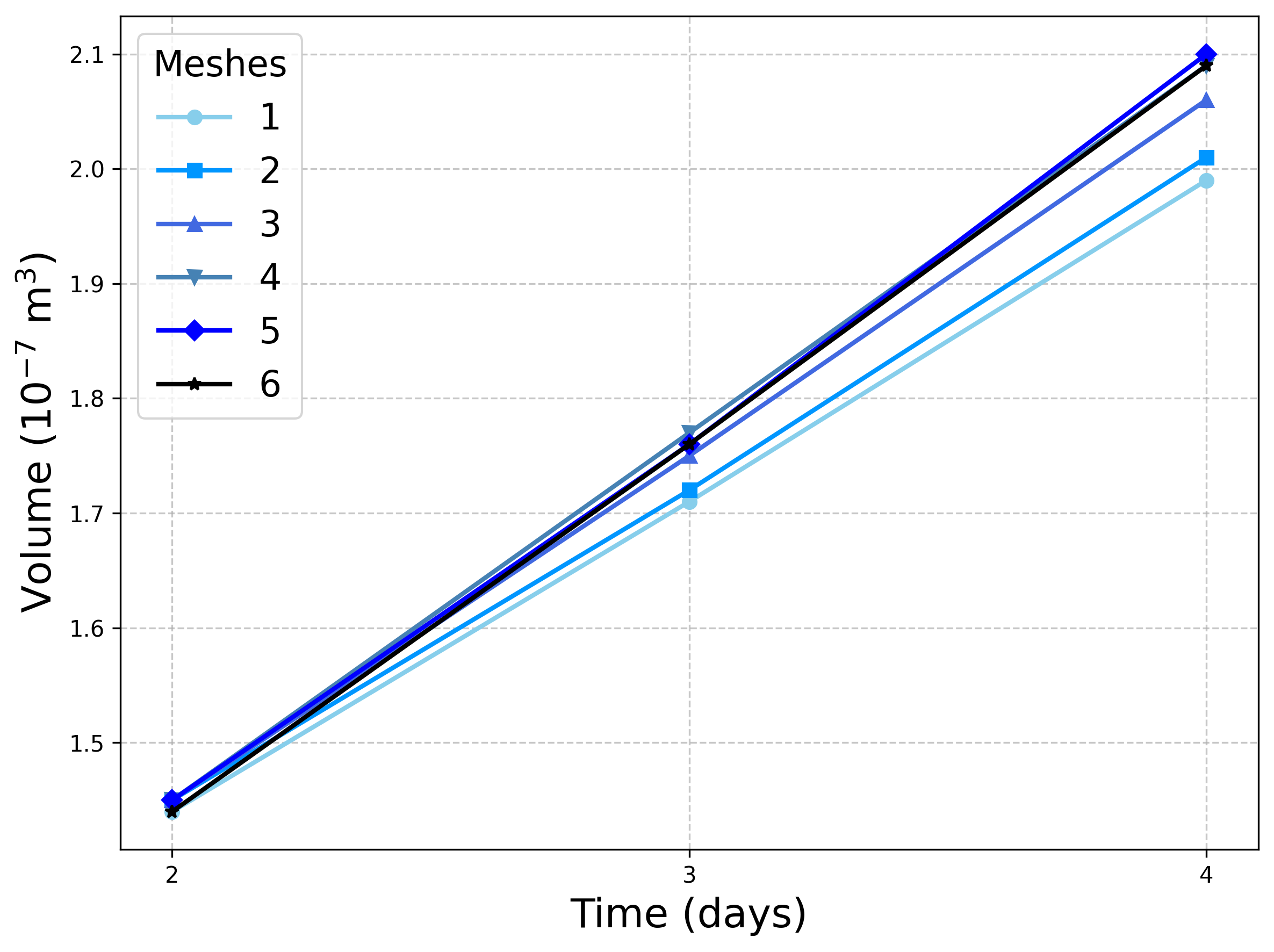}
    \caption{Tumor volumes versus time for rat 1}
    \label{sensi1b}
  \end{subfigure}
  \hfill
  \begin{subfigure}{0.43\textwidth}
    \includegraphics[width=\linewidth]{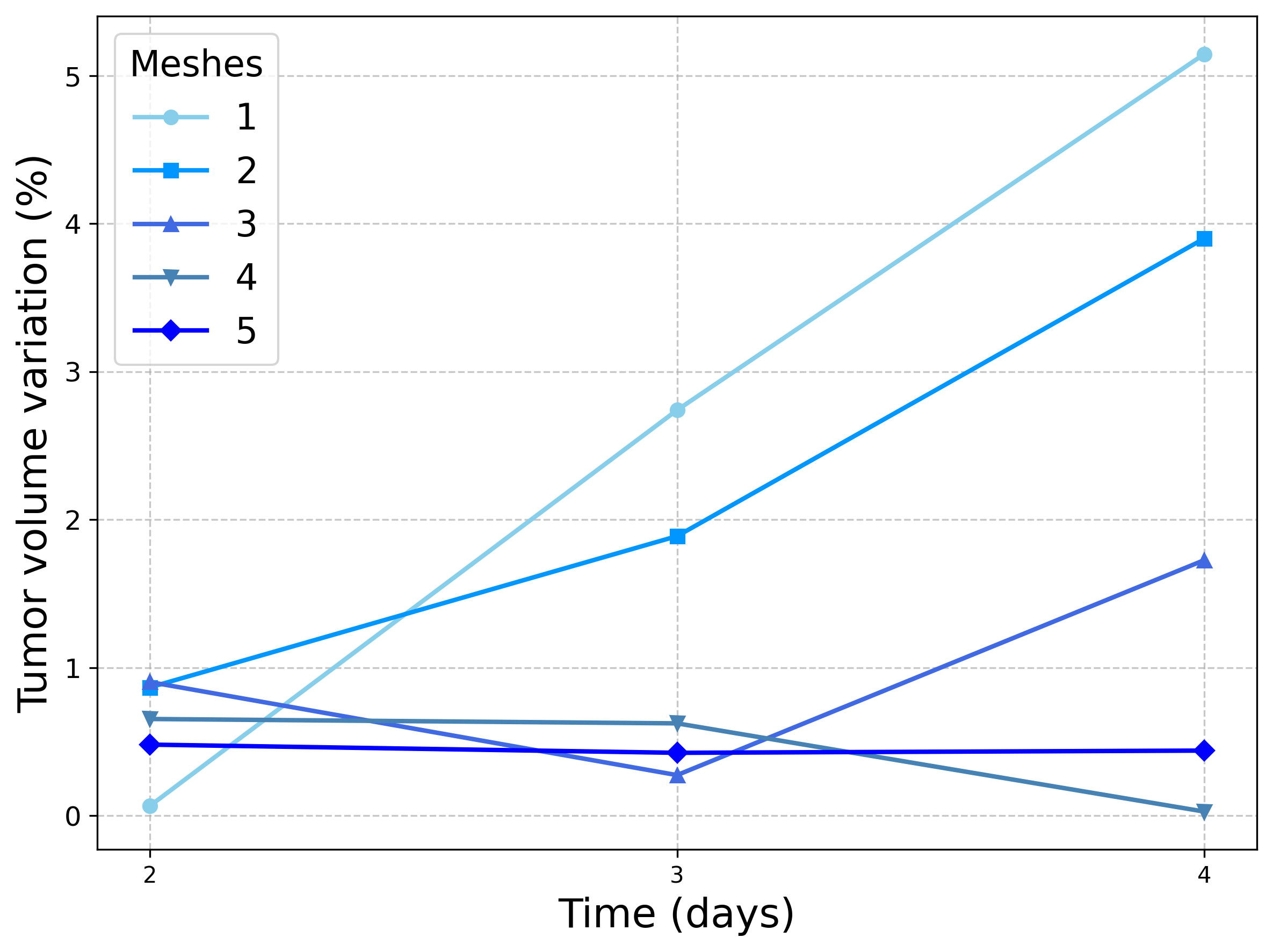}
    \caption{Tumor volume variances versus time for rat 1}
    \label{sensi1c}
  \end{subfigure}
  \vspace{0.3cm}
  \begin{subfigure}{0.8\textwidth}
    \includegraphics[width=\linewidth]{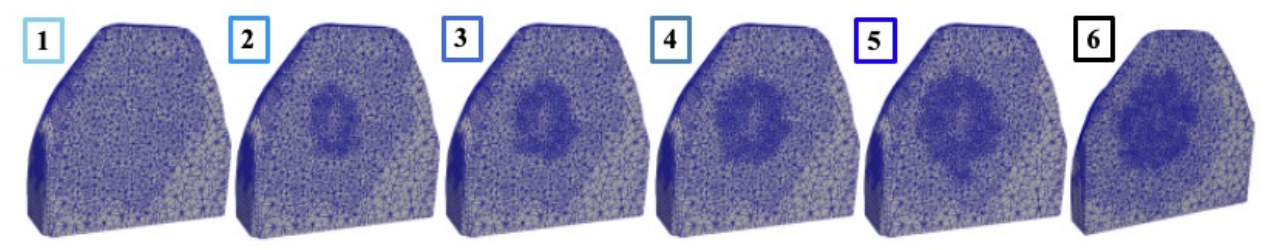}
    \caption{Meshes with different levels of refinement around the tumor for rat 2}
    \label{sensi2a}
  \end{subfigure}
  \hfill
  \begin{subfigure}{0.43\textwidth}
    \includegraphics[width=\linewidth]{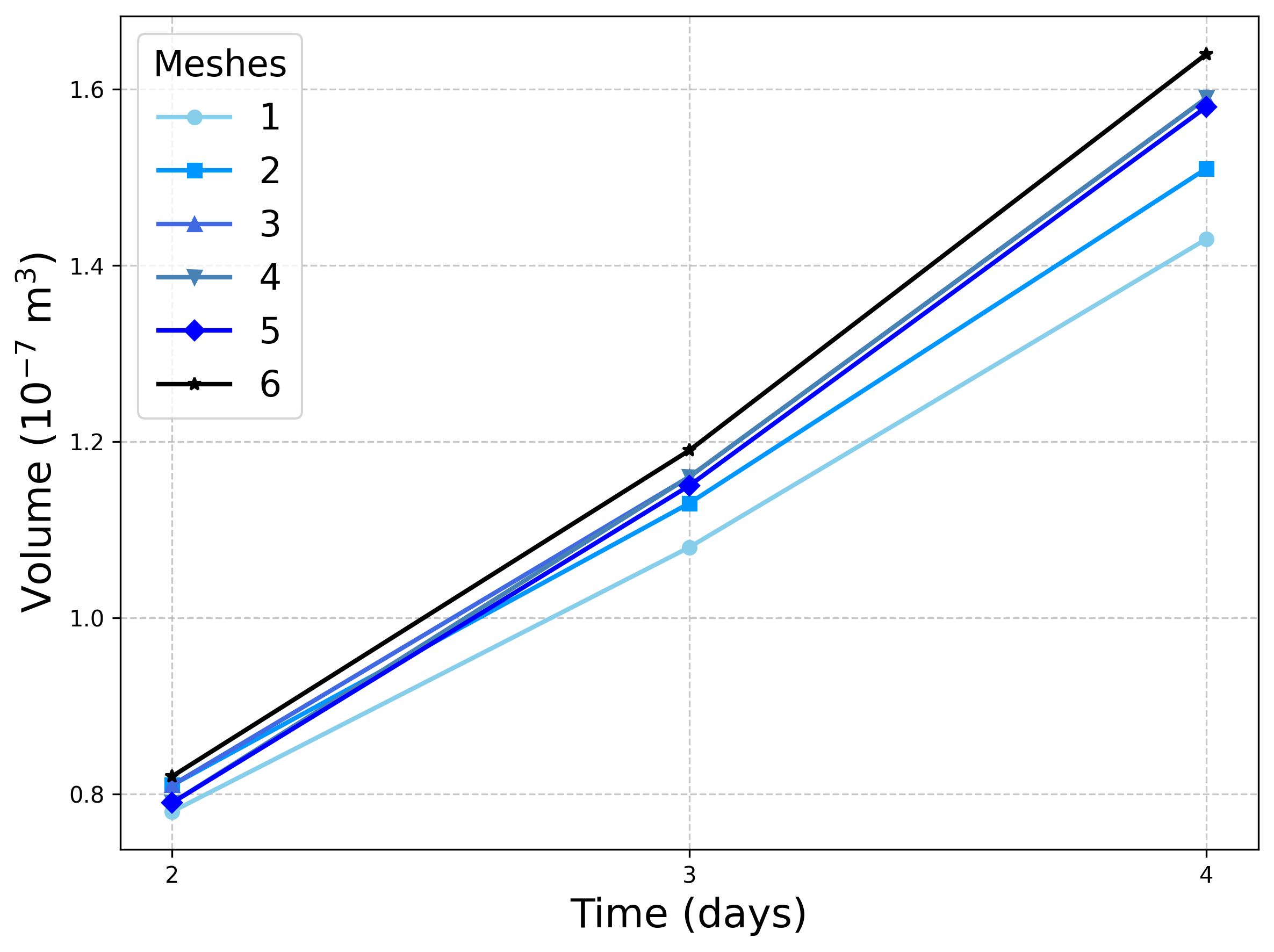}
    \caption{Tumor volumes versus time for rat 2}
    \label{sensi2b}
  \end{subfigure}
  \hfill
  \begin{subfigure}{0.43\textwidth}
    \includegraphics[width=\linewidth]{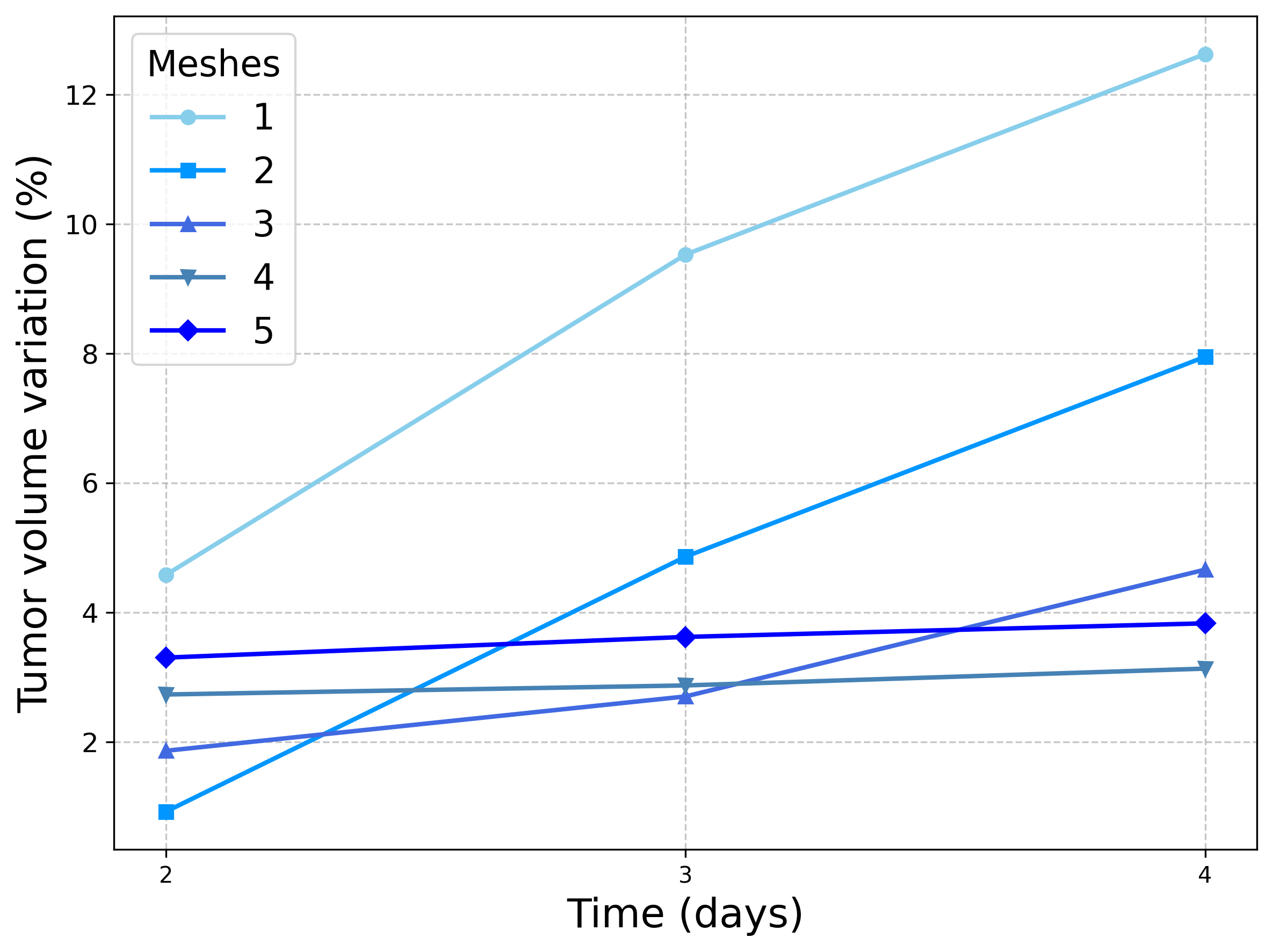}
    \caption{Tumor volume variances versus time for rat 2}
    \label{sensi2c}
  \end{subfigure}
  \caption{ Mesh sensitivity analysis for rats 1 and 2. For each animal, six meshes with increasing level of refinement were investigated. Mesh 6 was the reference mesh against which tumor volume errors at days 2, 3, and 4 were computed. Mesh 3 was the chosen mesh for the computations in the rest of our study as it best balanced accuracy and efficiency (see number of elements in each mesh in Table~\ref{elem}. Continues on next page.}
  \label{mainfig}
\end{figure}

\begin{figure}[p]
  \centering
  \begin{subfigure}{0.8\textwidth}
    \includegraphics[width=\linewidth]{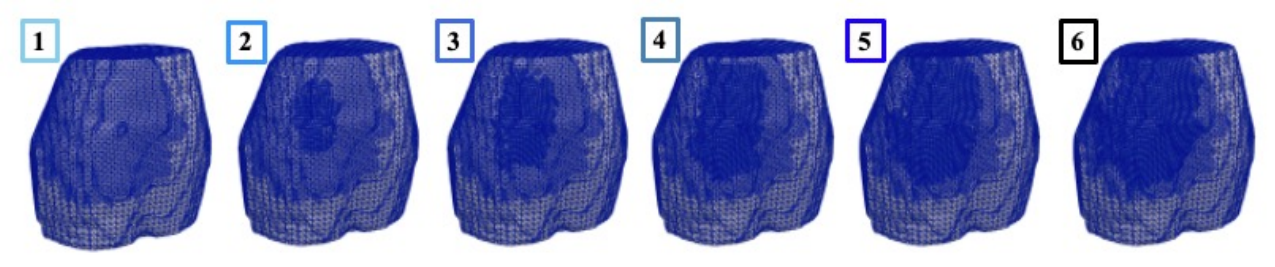}
    \caption{6 Meshes with different levels of refinement around the tumor for rat 3}
    \label{sensi3a}
  \end{subfigure}
  \hfill
  \begin{subfigure}{0.43\textwidth}
    \includegraphics[width=\linewidth]{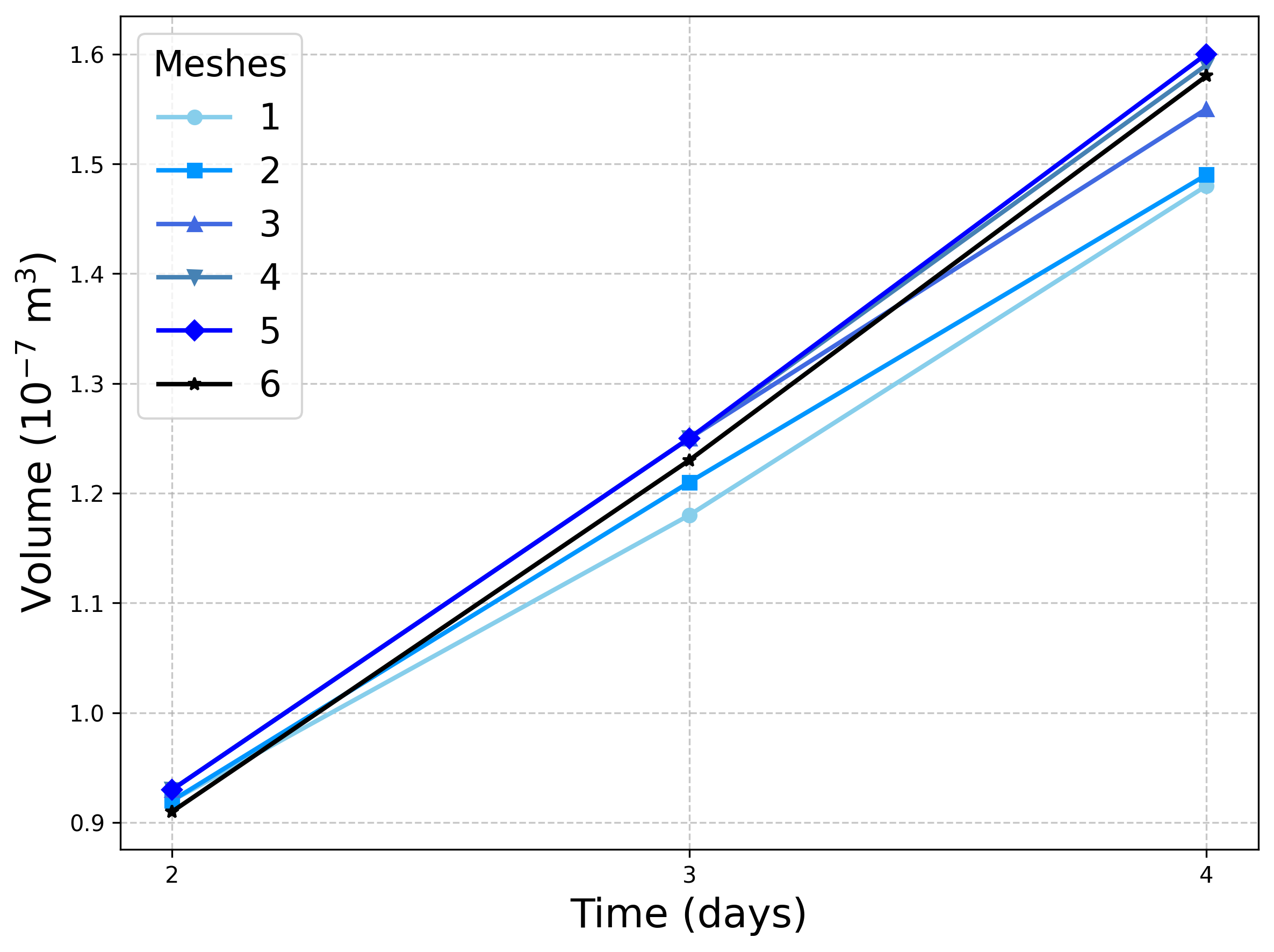}
    \caption{Tumor volumes versus time for rat 3}
    \label{sensi3b}
  \end{subfigure}
  \hfill
  \begin{subfigure}{0.43\textwidth}
    \includegraphics[width=\linewidth]{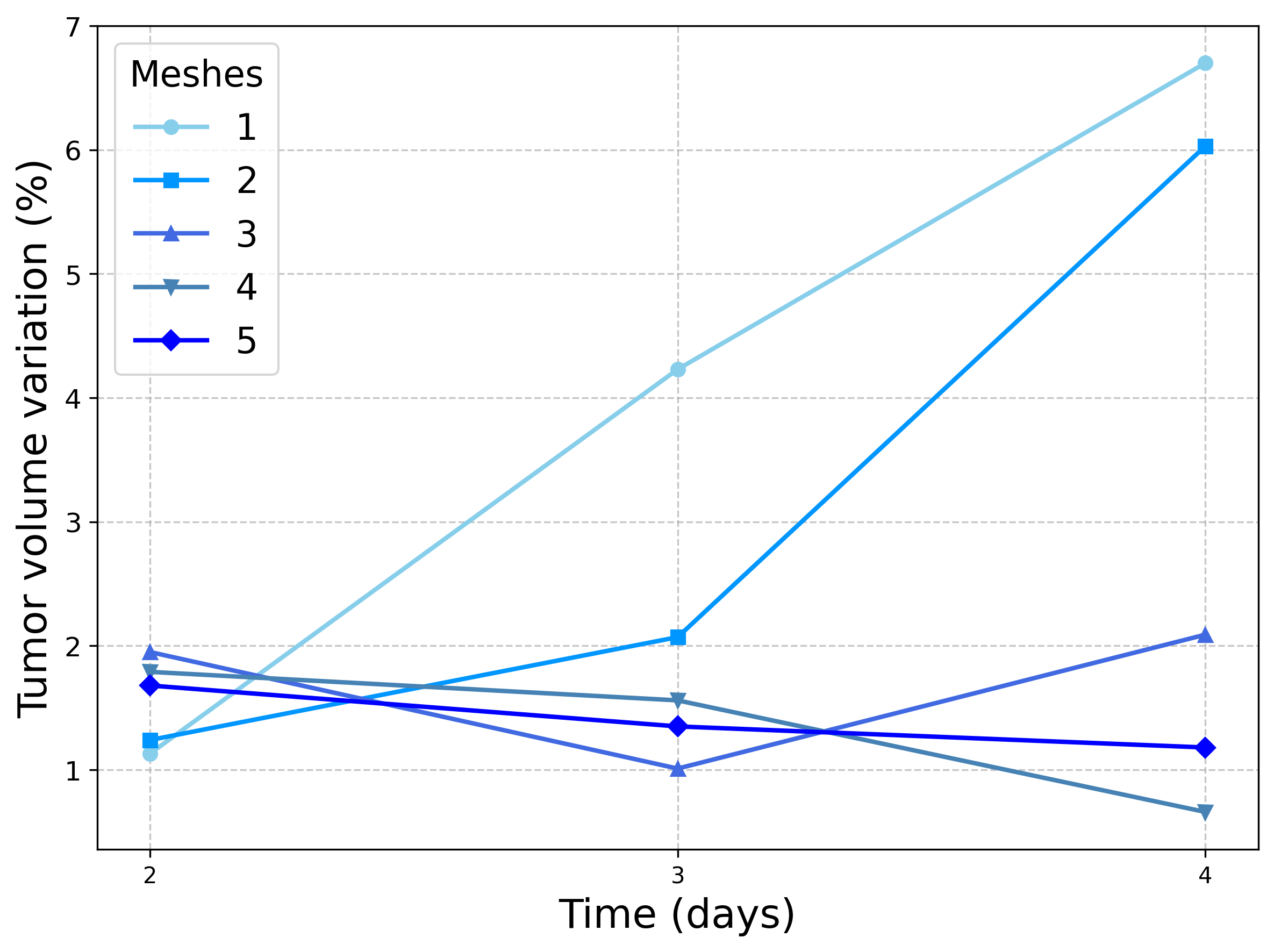}
    \caption{Tumor volume variances versus time for rat 3}
    \label{sensi3c}
  \end{subfigure}
  \vspace{0.3cm}
  \begin{subfigure}{0.8\textwidth}
    \includegraphics[width=\linewidth]{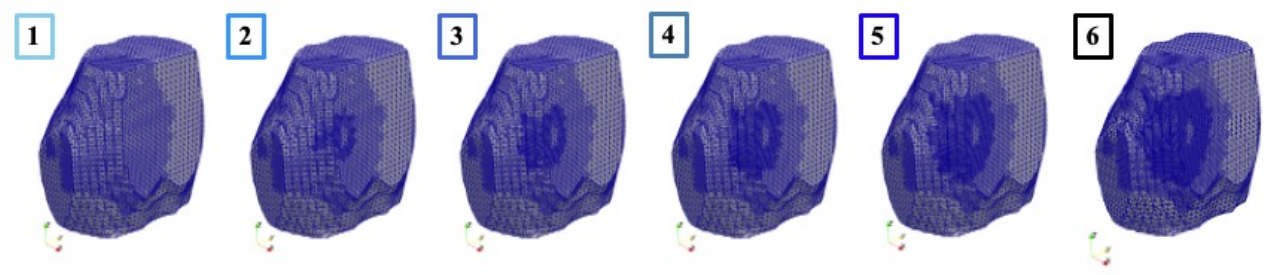}
    \caption{6 Meshes with different levels of refinement around the tumor for rat 4}
    \label{sensi4a}
  \end{subfigure}
  \hfill
  \begin{subfigure}{0.43\textwidth}
    \includegraphics[width=\linewidth]{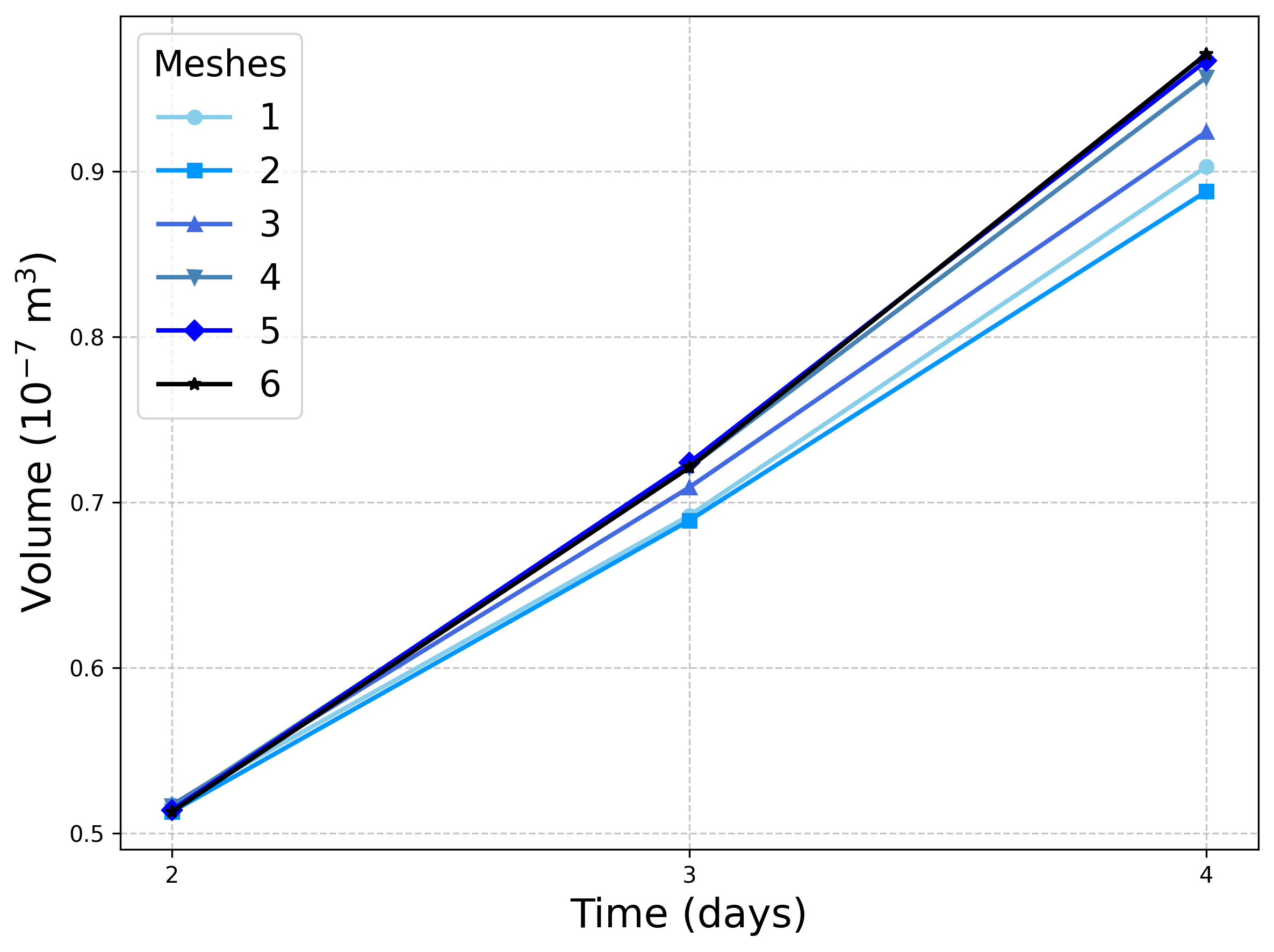}
    \caption{Tumor volumes versus time for rat 4}
    \label{sensi4b}
  \end{subfigure}
  \hfill
  \begin{subfigure}{0.43\textwidth}
    \includegraphics[width=\linewidth]{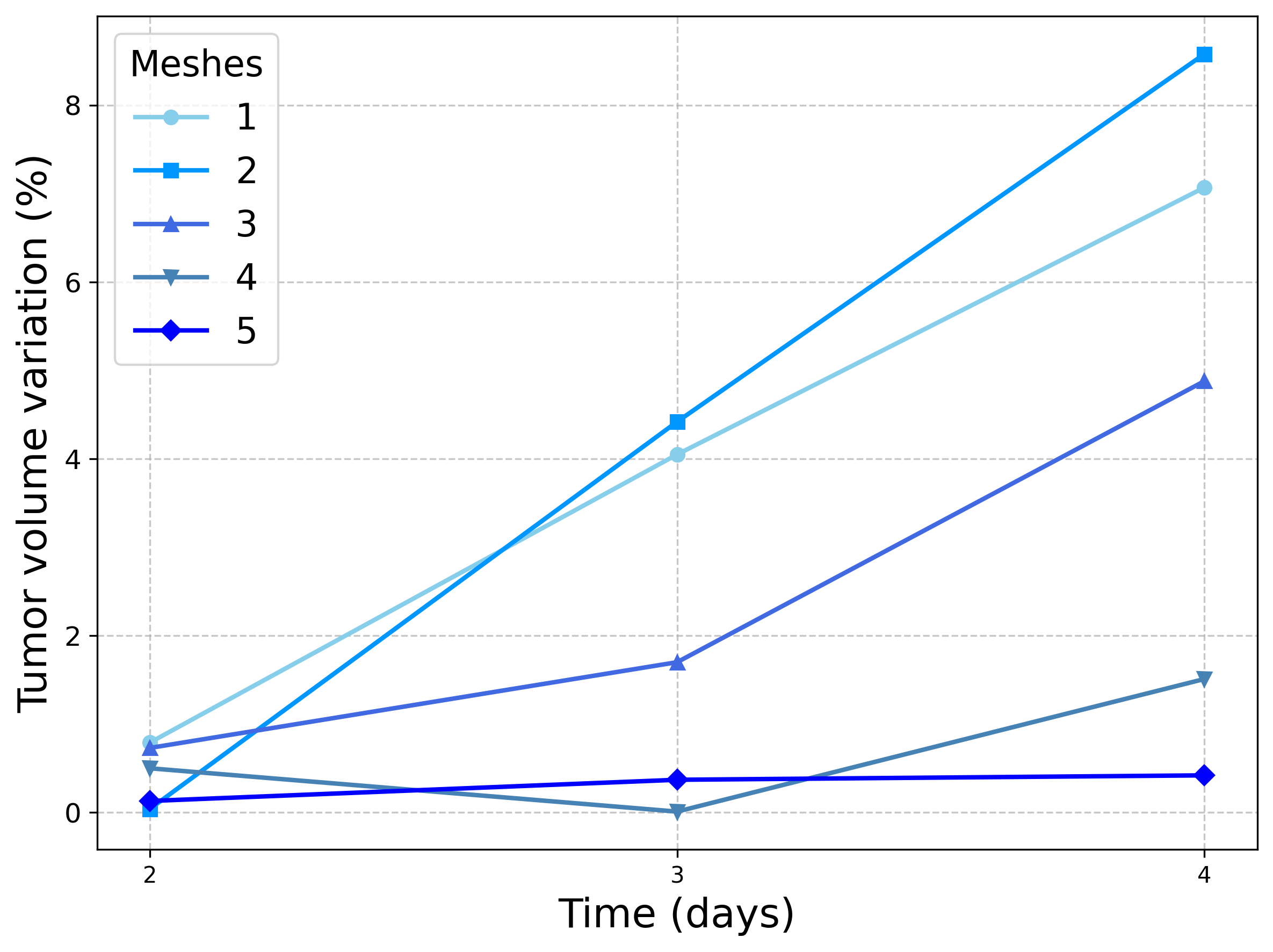}
    \caption{Tumor volume variances versus time for rat 4}
    \label{sensi4c}
  \end{subfigure}
  \caption{Mesh sensitivity analysis for rats 3 and 4. For each animal, six meshes with increasing level of refinement were investigated. Mesh 6 was the reference mesh against which tumor volume errors at days 2, 3, and 4 were computed. Mesh 3 was the chosen mesh for the computations in the rest of our study as it best balanced accuracy and efficiency (see number of elements in each mesh in Table~\ref{elem}.}
\end{figure}

We first analyzed the numerical accuracy and computational cost of the six candidate meshes generated for each rat in this study (see Fig.~\ref{sensi1a},~\ref{sensi2a},~\ref{sensi3a} and~\ref{sensi4a}). The number of elements for each mesh and each rat are shown in Table~\ref{elem}. Meshes 1 and 2  were coarser and generally showed higher tumor volume error  than other meshes for all animals. For example, mesh 1 resulted in $9.53 \%$ error at day 3 and 12.6$\%$ at day 4 for rat 2 (see Fig.~\ref{sensi2b} and~\ref{sensi2c}), representing the upper bound for tumor volume error in the mesh sensitivity study. The computation time of mesh 3 demonstrated significant time savings compared to more refined meshes  accross all four rats. For example, for rat 4, mesh 3 was $23.9 \%$, $43.5 \%$ and $57 \%$ faster than meshes 4, 5, and 6 (35.22, 86.37 and 113.47 minutes faster, respectively). Additionally, mesh 3 provided the best balance between accuracy and computational efficiency across animals, with less than $3\%$ error by day 3 and $5\%$ by day 4 across all rats  (see Figs.~\ref{sensi1b},~\ref{sensi2b},~\ref{sensi3b},~\ref{sensi4b}). Thus, we leveraged mesh 3 for each rat to proceed with the rest of the calculations in this work. 

\begin{table}[ht]
\centering
 \begin{tabular*}{\textwidth}{@{\extracolsep{\fill}}ccccccc}

\toprule
Rat & Mesh 1 & Mesh 2 & Mesh 3 & Mesh 4 & Mesh 5 & Mesh 6 \\
\midrule
Rat 1 & 127,291 & 198,312 & 229,933 &264,687 & 305,432 & 352,028\\
Rat 2 & 151,421 & 186,654 & 214,386 &244,125 &290,970 & 350,797 \\
Rat 3 & 102,282 & 158,024 &190,093 & 230,281 & 270,156 & 316,380 \\
Rat 4 & 90,897 & 120,077 & 134,987 & 154,890 & 177,328 & 204,256 \\
\bottomrule
\end{tabular*}
\caption{Number of elements numbers across the six meshes with increasing level of refinement for rats 1 to 4 }\label{elem}
\end{table}

\subsection{Variance-based sensitivity analysis}

\begin{figure}
    \centering
    \includegraphics[width=0.5\columnwidth]{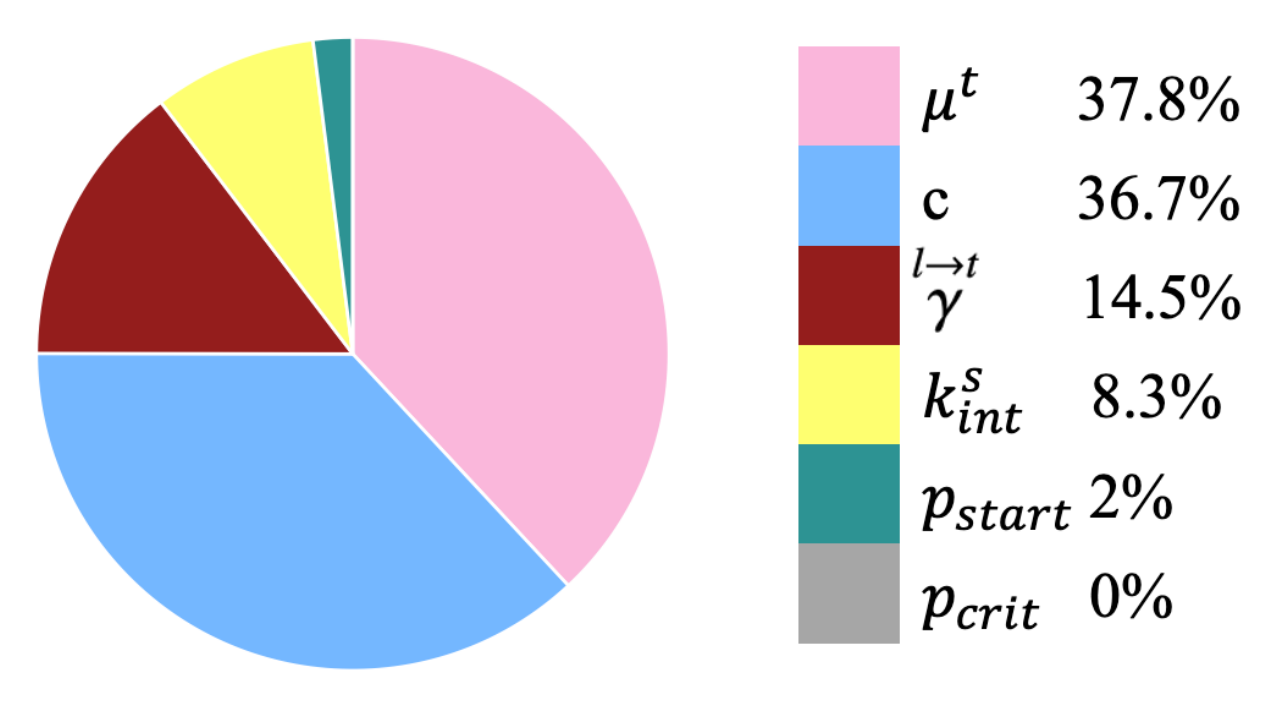}
    \caption{Parameter variance-based sensitivity analysis. The tumor growth rate ($\mathop \gamma\limits^{l \to t}$), the intrinsic permeability ($k^s_{int}$), the pressure-saturation parameter, and the tumor cell dynamic viscosity ($\mu^t$) contributed to more than $90 \%$ of the variance of the model solution and, therefore, they were considered influential. Conversely, the starting and critical solid pressure thresholds controlling the mechanical inhibition of tumor proliferation (see Eq.~\eqref{chap5.29}) were found to be not influential in this study.}
    \label{Sobol}
\end{figure}

Fig.~\ref{Sobol} summarizes the results of the variance-based analysis of the six parameters that we identified as candidates to drive tumor dynamics in our poroelastic modeling framework. We found that variations in the tumor growth rate ($\mathop \gamma\limits_{} \limits^{l \to t}$), the intrinsic permeability ($k^{s}_{int}$), the pressure-saturation parameter, and the tumor cell dynamic viscosity ($\mu^t$) contribute to more than $90 \%$ of the variance in model solution dynamics. Conversely, the starting and critical solid pressures controlling the mechanical inhibition of tumor proliferation (see Eq.~\eqref{chap5.29}) contributed less than $2 \%$ to the total model solution variance. Since these parameters were not significant they were chosen such that the mechanical inhibition function $\mathcal{H}(p^{s}, p_{\text{start}}, p_{\text{crit}})=1$ for all glioma simulations in this work.

\begin{table}
\centering
 \begin{tabular*}{\textwidth}{@{\extracolsep{\fill}}c | c c c c |c c c c@{}}

    \toprule
      set & c & $\mathop \gamma\limits_{} \limits^{l \to t}$ & $k^{s}_{int}$ ($\alpha_k$) & $\mu^{t}$ &  Rat 1 & Rat 2 & Rat 3 & Rat 4 \\
        \midrule
        \midrule
        1 & 500  & $5.29\times 10^{-5}$ & 1 & 40 & 6.73 & 17.21 &12.16 & 10\\
        \midrule
         2 & 500  & $5.29 \times 10^{-5}$ & 1 & 70 & 6.52 & 16.82 & 18.43 & 14.08 \\
         \midrule
         3 &500  & $5.29\times 10^{-5}$ & 1 & 80 & 7.43 & 17.68 & 19.98 & 16.27\\
         \midrule
        4 &500  & $5.29\times 10^{-5}$ & 1 & 90 & 8.47 & 18.61 &21.29 & 17.88\\
        \midrule
         5& 500 & $5.29\times 10^{-6}$  & 0.1 & 80 &  22.54 & 34.65 & 34.10 & 34.55 \\
         \midrule
         6&500   & $5.29\times 10^{-6}$ & 0.35 & 80  & 16.9 & 27.7 & 29.47 & 28.28 \\
         \midrule
          7& 500 & $5.29\times 10^{-6}$ & 0.55 & 80 & 13.5 & 23.7 & 26.45 & 24.55 \\
        \midrule
          8& 1000 &$5.29\times 10^{-5}$ & 0.55 & 70 & 6.07 & 16.18 & 17.27 & 12.93 \\
        \midrule
          9& 1000 & $5.29\times 10^{-4}$ & 0.55 & 70 &  6.05 & 17.35 & \bf{11.83} & 9.49 \\
        \midrule
          10& 1000 & $5.29\times 10^{-3}$& 0.65 & 40 &  -  & - & - & -    \\
        \midrule
          11& 500 &$5.29\times 10^{-3}$ & 0.65 & 40 & -  & - & - & -  \\
        \midrule
          12& 1000 & $5.29\times 10^{-6}$ & 0.1 & 40 &  16 & 26.64 & 28.67 & 27.27  \\
        \midrule
          13& 500 & $5.29\times 10^{-3}$ & 0.65 & 40 & -  & - & - & -   \\
         \midrule
          14& 500 & $5.29\times 10^{-4}$& 0.55 & 70 & 7.79  & 16.77  & 18.92 & 14.44\\
        \midrule
          15& 500 & $5.29\times 10^{-4}$& 0.55 & 40 & 5.37  & 16.03  & 12.73 & \bf{8.78}\\
        \midrule
          16& 500 & $5.29\times 10^{-4}$ & 0.45 & 40 & 5.48  & 15.32 & 14.76 & 9.88\\
        \midrule
          17& 500 & $5.29\times 10^{-5}$ & 0.35 & 80 & 16.49  & 27.09 & 28.89 & 27.36\\
        \midrule
         18& 1000 &$5.29\times 10^{-4}$ & 0.55 & 90 & \bf{5.28}  & \bf{14.99} & 13.95 & 9.14\\

    \bottomrule

  \end{tabular*}

\caption{Parameter combinations used to identify an appropriate initial guess for model calibration using the GN method along with their corresponding GN error calculated through the loss function, $L(\varepsilon^t,\theta)$, at day 3 for all rats. GN errors in bold identify the optimal initial parameter combination for each rat. Computations diverged for all rats with parameter combinations 10, 11 and 13.}
\label{paramspace}
\end{table}

\subsection{Parameter calibration}
\label{Parameter_calibration}
To calculate predictions of glioma growth for each individual rat, we calibrate the four relevant parameters identified in the variance-based sensitivity analysis to the longitudinal MRI data measured from each animal up to day 3 using the GN method. To avoid landing on a parameter combination yielding a local minimum of the GN error calculated through the GN loss function, we examined a set of 18 parameter combinations spanning the range of values in the literature \cite{urcun2021digital,urcun2023non,sciume2013multiphase}. We solved the model up to day 3 and tabulated the GN error for each parameter combination for each rat. Table~\ref{paramspace} lists the 18 parameter combinations and the corresponding GN error obtained in each rat. For the subsequent animal-specific parameter calibration, we chose as initial guess the parameter combination achieving the lowest GN error for each rat. 
Applying the GN method to further refine the parameter values for each rat resulted in only marginal error reductions across all subjects: $0.12$ \% for Rat 1, $0.31$ \% for Rat 2, $0.18$ \% for Rat 3, and $0.11$ \% for Rat 4. All four parameters increased between the initial and optimized states as shown in Table~\ref{opti}.

\begin{table}
\centering
 \begin{tabular*}{\textwidth}{@{\extracolsep{\fill}}c|c|c c c c|c c c@{}}
    \toprule
      Rat & state & c & $\mathop \gamma\limits_{} \limits^{l \to t}$ & $k^{s}_{int}$ ($\alpha_k$) & $\mu^{t}$ &  \makecell{GN \\($\%$)} & \makecell{NIV \\($\%$)} & DSC \\
        \midrule
        1 & initial& 1000  & $5.290\times 10^{-4}$ & 0.550 & 90 & 5.60  & 14.52 & 0.92\\

        1 & optimized &1012.32  & $5.366\times 10^{-4}$ & 0.569 & 99 & 5.48  & 14.29 & 0.93 \\

         \midrule
         2 & initial& 1000  & $5.29\times 10^{-4}$ & 0.550 & 90 & 15.02 & 35.76 & 0.82 \\

         2 & optimized& 1070.27  & $5.24\times 10^{-4}$ & 0.544 & 99.86 & 14.71 & 35.01 & 0.82 \\

        \midrule
         3 & initial&1000  & $5.29\times 10^{-4}$ & 0.550 & 70 & 11.8 & 28.03 & 0.82 \\

         3 & optimized & 1418.54  & $5.51\times 10^{-4}$ & 0.568 & 98.22 & 11.62 & 27.74 & 0.82 \\

        \midrule
        4 & initial&500  & $5.29\times 10^{-4}$ & 0.550 & 40 & 8.84 & 24.49 & 0.80 \\

        4 & optimized&576.9  & $5.46\times 10^{-4}$ & 0.620 & 53.55 & 8.73 & 24.36 &0.80 \\

    \bottomrule
  \end{tabular*}
\caption{Initial and optimal parameter sets along with corresponding model calibration errors at day 3 for all subjects.}\label{opti}
\end{table}

\subsection{Tumor volume predictions}
\begin{figure}[!htp]
  \centering
  \begin{subfigure}{0.18\textwidth}
    \includegraphics[width=\linewidth]{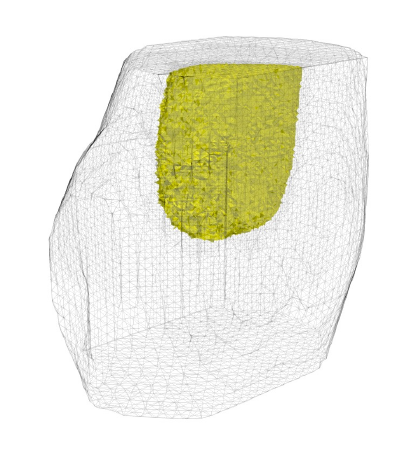}
    \caption*{Initial tumor}
  \end{subfigure}
  \begin{subfigure}{0.18\textwidth}
    \includegraphics[width=\linewidth]{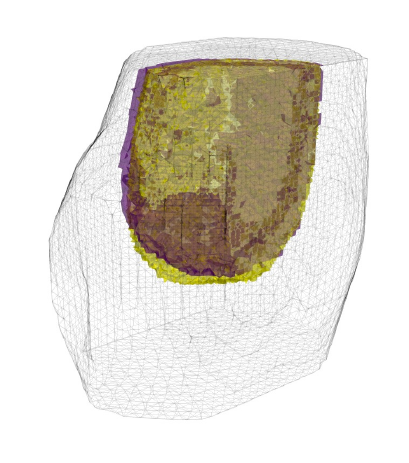}
    \caption*{Day 3. DSC = 0.92}
  \end{subfigure}
  \begin{subfigure}{0.18\textwidth}
    \includegraphics[width=\linewidth]{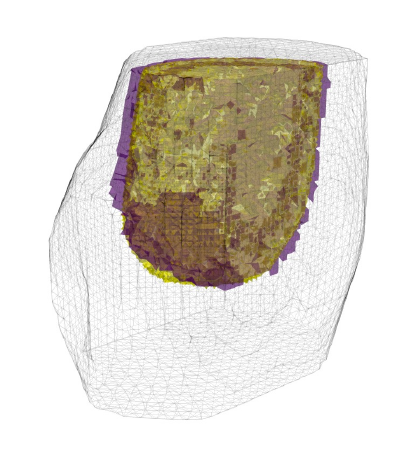}
    \caption*{Day 4. DSC = 0.93}
  \end{subfigure}
  \begin{subfigure}{0.18\textwidth}
    \includegraphics[width=\linewidth]{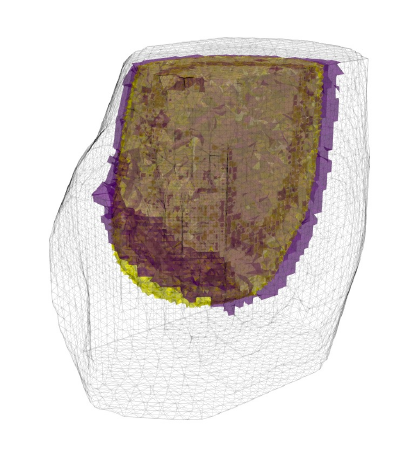}
    \caption*{Day 6. DSC = 0.90}
  \end{subfigure}\\[1ex]
  \begin{subfigure}{\textwidth}
    \centering
    \caption{Rat 1}
    \label{modelrat1}
  \end{subfigure}\\[2ex]

  \begin{subfigure}{0.18\textwidth}
    \includegraphics[width=\linewidth]{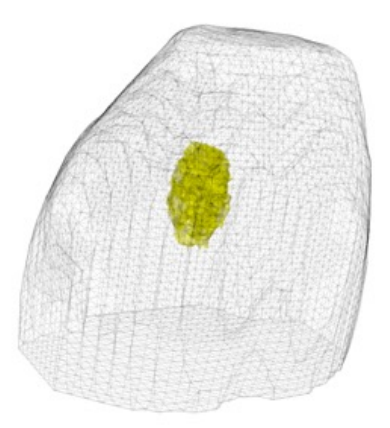}
    \caption*{Initial tumor}
  \end{subfigure}
  \begin{subfigure}{0.18\textwidth}
    \includegraphics[width=\linewidth]{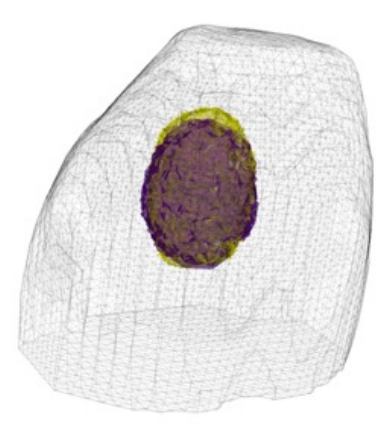}
    \caption*{Day 3. DSC = 0.82}
  \end{subfigure}
  \begin{subfigure}{0.18\textwidth}
    \includegraphics[width=\linewidth]{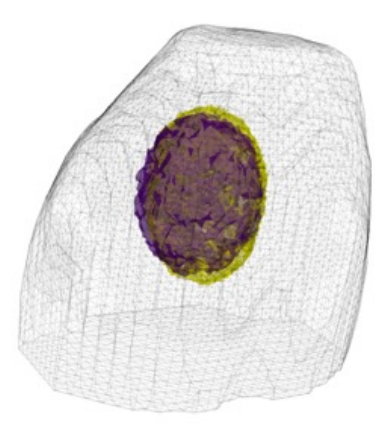}
    \caption*{Day 4. DSC = 0.81}
  \end{subfigure}
  \begin{subfigure}{0.18\textwidth}
    \includegraphics[width=\linewidth]{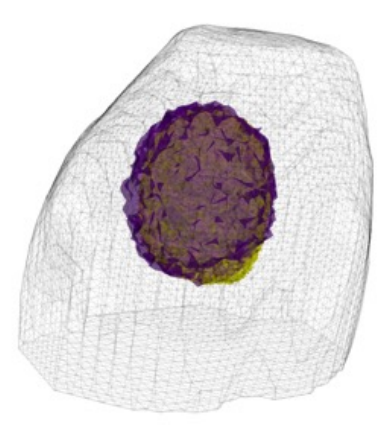}
    \caption*{Day 6. DSC = 0.75}
  \end{subfigure}\\[1ex]
  \begin{subfigure}{\textwidth}
    \centering
    \caption{Rat 2}
    \label{modelrat2}
  \end{subfigure}\\[2ex]

  \begin{subfigure}{0.18\textwidth}
    \includegraphics[width=\linewidth]{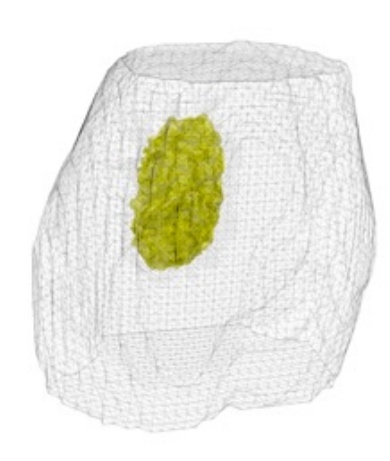}
    \caption*{Initial tumor}
  \end{subfigure}
  \begin{subfigure}{0.18\textwidth}
    \includegraphics[width=\linewidth]{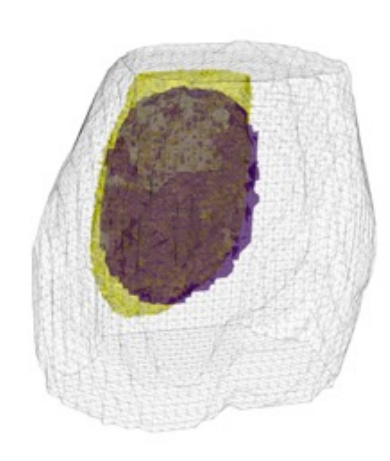}
    \caption*{Day 3. DSC = 0.82}
  \end{subfigure}
  \begin{subfigure}{0.18\textwidth}
    \includegraphics[width=\linewidth]{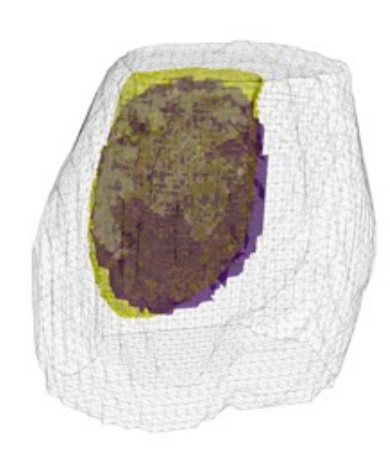}
    \caption*{Day 4. DSC = 0.87}
  \end{subfigure}
  \begin{subfigure}{0.18\textwidth}
    \includegraphics[width=\linewidth]{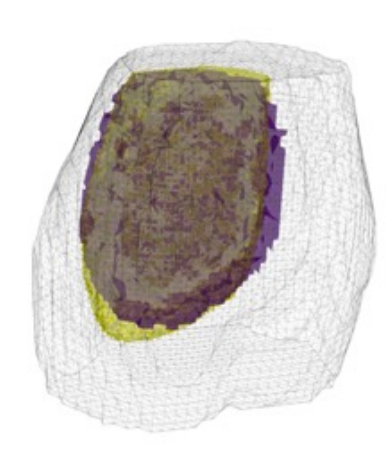}
    \caption*{Day 6. DSC = 0.87}
  \end{subfigure}\\[1ex]
  \begin{subfigure}{\textwidth}
    \centering
    \caption{Rat 3}
    \label{modelrat3}
  \end{subfigure}\\[2ex]

  \begin{subfigure}{0.18\textwidth}
    \includegraphics[width=\linewidth]{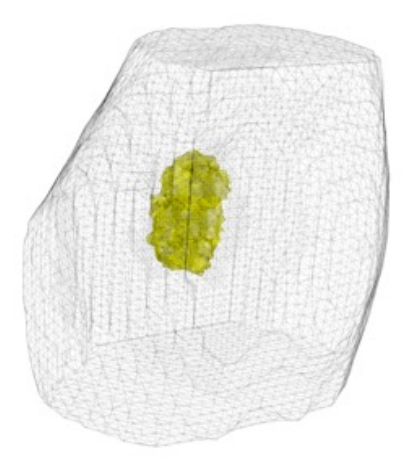}
    \caption*{Initial tumor}
  \end{subfigure}
  \begin{subfigure}{0.18\textwidth}
    \includegraphics[width=\linewidth]{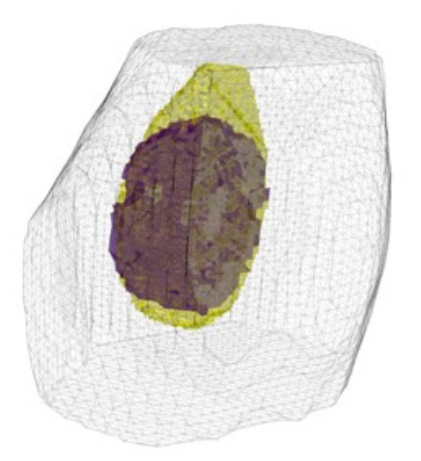}
    \caption*{Day 3. DSC = 0.80}
  \end{subfigure}
  \begin{subfigure}{0.18\textwidth}
    \includegraphics[width=\linewidth]{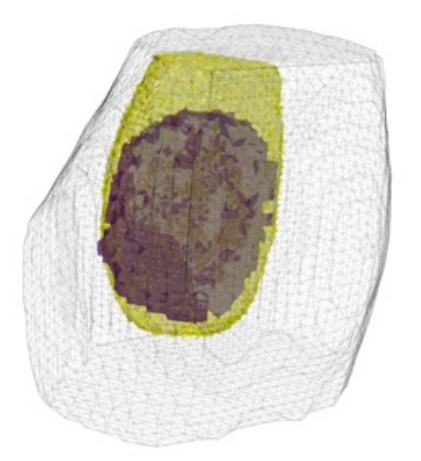}
    \caption*{Day 4. DSC = 0.69}
  \end{subfigure}
  \begin{subfigure}{0.18\textwidth}
    \includegraphics[width=\linewidth]{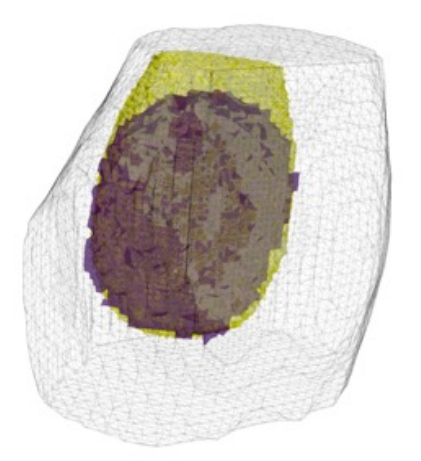}
    \caption*{Day 6. DSC = 0.77}
  \end{subfigure}\\[1ex]
  \begin{subfigure}{\textwidth}
    \centering
    \caption{Rat 4}
    \label{modelrat4}
  \end{subfigure}

  \caption{
    Front views of predicted (purple) and observed (yellow) tumor evolution for all four rats. Each row shows tumor shape at initial, calibration, and both prediction times, with Dice Similarity Coefficient (DSC) values noted. DSC values ranged from 0.69 (rat 4) to 0.94 (rat 1).
  }
  \label{fig:allrats}
\end{figure}

\begin{figure}[H]
  \centering
  \begin{subfigure}{0.22\textwidth}
    \includegraphics[width=\linewidth]{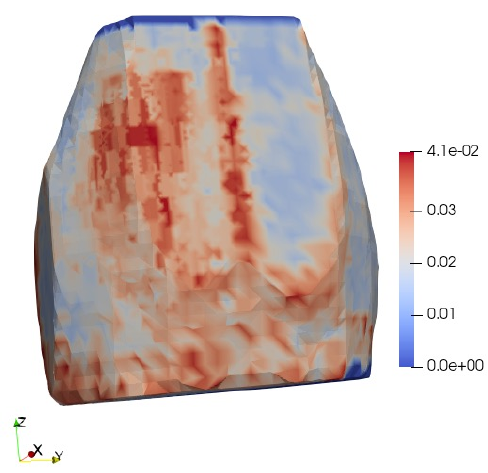}
    \caption{Vascularization field $\varepsilon^s\omega^{bs}$ for rat 4}
    \label{}
  \end{subfigure}
  \hfill
  \begin{subfigure}{0.22\textwidth}
    \includegraphics[width=\linewidth]{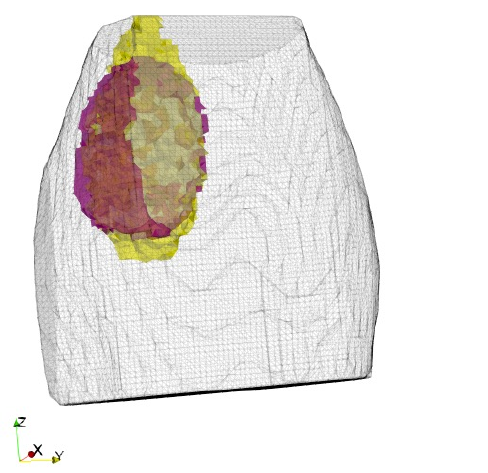}
    \caption{Tumor volume observation (yellow) and prediction (purple) at day 3}
    \label{}
  \end{subfigure}
  \hfill
  \begin{subfigure}{0.36\textwidth}
    \includegraphics[width=\linewidth]{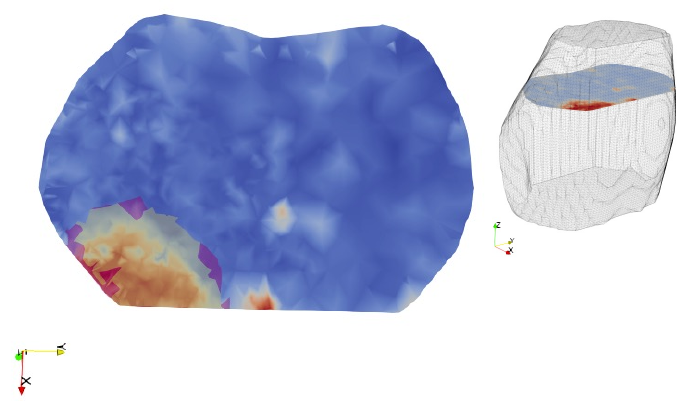}
    \caption{Axial slice of vascularization overlapped with predicted and observed tumor volumes}
    \label{}
  \end{subfigure}
  \caption{Comparison of tumor volume observation and prediction with the vascularization field at day 3 for rat 4. Tumor predictions seem to follow the external vascularization pattern. The observed tumor, however, also includes regions with low vascularization. 
  }
  \label{ratwbs}
\end{figure}

Table~\ref{pred} presents errors in predicting tumor volume on day 4 and day 6 for each rat, evaluated using NIV, DSC, and $e_{vol}$. On day 4, DSC values ranged from 0.69 to 0.93, with rat 1 showing the highest spatial overlap and rat 4 the lowest. NIV values varied widely, from 11.77 to 47.54, indicating differences in non-overlapping regions across subjects. Relative tumor volume errors on day 4 ranged from 4.73$\%$ to $36.03 \%$. On day 6, DSC values remained relatively high (0.74 to 0.90), while NIV values ranged from 17.99$\%$ to 50.61$\%$ and relative volume errors from 15.96$\%$ to 30.74$\%$. 
Overall, rat 1 showed the best model-data agreement in terms of tumor volume estimation at both prediction times with DSC values higher than $0.90$ while rat 4 exhibited the furthest results with a DSC of $0.69$ at day 4. Tumor volume prediction against observations are shown in Figs.~\ref{modelrat1},~\ref{modelrat2},~\ref{modelrat3} and~\ref{modelrat4} for all rats. 
Additionally, Fig.~\ref{ratwbs} provides further insight in the underlying causes of the quantitative prediction errors in Table~\ref{pred}, particularly for rat 4, which showed the poorest performance (DSC=0.69, $e_{vol}$=36.03 \% on day 4). Fig.~\ref{ratwbs} overlays initial vascularization with tumor observations and predictions at day 4, revealing that model predictions closely track highly vascularized regions, yet the observed tumor also extends into regions with relatively low vascularization. This observation indicates that vascularization alone does not fully constrain the spatial growth of the tumor, and that further mechanisms should be investigated.

\begin{table}
 \begin{tabular*}{\textwidth}{@{\extracolsep{\fill}}c| c c c|c c c@{}}
\toprule
\multirow{2}{*}{Rat} 
  & \multicolumn{3}{c|}{Day 4} 
  & \multicolumn{3}{c}{Day 6} \\
\cmidrule(lr){2-4} \cmidrule(l){5-7}
  & NIV ($\%$) & DSC & $e_{vol} ($\%$)$& NIV ($\%$) & DSC & $e_{vol}$ ($\%$)\\
\midrule
  1 & 11.77 & 0.93 & 11.09 & 17.99 & 0.90 & 15.96 \\
  2 & 36.66 & 0.81 &  4.73 & 50.61 & 0.75 & 30.74 \\
  3 & 22.23 & 0.87 & 15.68 & 25.24 & 0.87 & 15.36 \\
  4 & 47.54 & 0.69 & 36.03 & 34.69 & 0.77 & 19.05 \\
\bottomrule
\end{tabular*}
\caption{Errors on tumor volume at days 4 and 6 (i.e., first and second prediction times).}
\label{pred}
\end{table}
\newpage

\subsection{Analysis of model predictions}
\subsubsection{Solid displacement}
\begin{figure}[H]
  \centering
  \begin{subfigure}{0.22\textwidth}
    \includegraphics[width=\linewidth]{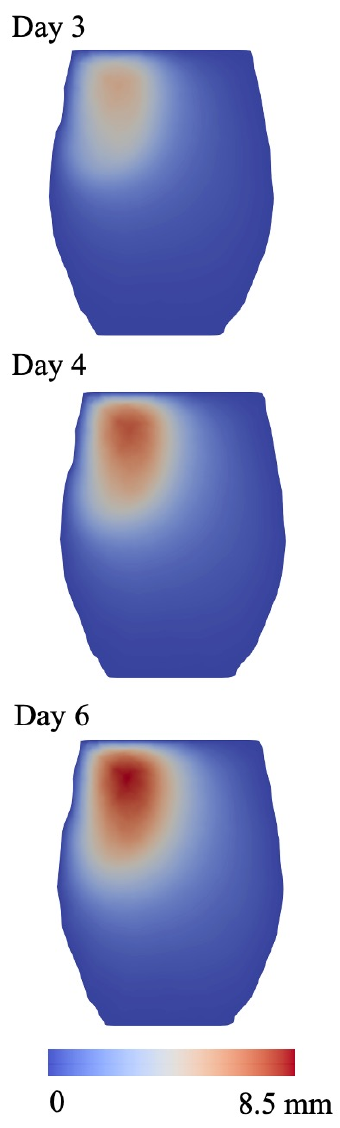}
    \caption{$\mathbf{u^s}$ for rat 1}
    \label{fig:dispdiff:a}
  \end{subfigure}
  \hfill
  \begin{subfigure}{0.22\textwidth}
    \includegraphics[width=\linewidth]{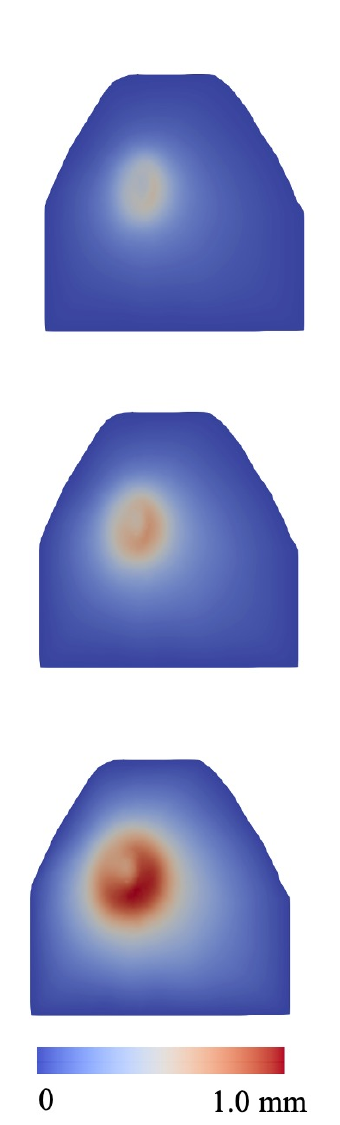}
    \caption{$\mathbf{u^s}$ for rat 2}
    \label{fig:dispdiff:b}
  \end{subfigure}
  \hfill
  \begin{subfigure}{0.22\textwidth}
    \includegraphics[width=\linewidth]{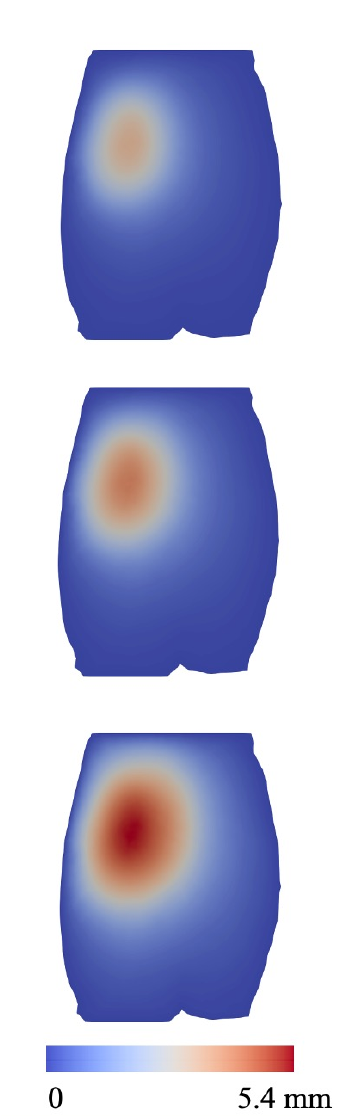}
    \caption{$\mathbf{u^s}$ for rat 3}
    \label{fig:dispdiff:c}
  \end{subfigure}
  \hfill
  \begin{subfigure}{0.22\textwidth}
    \includegraphics[width=\linewidth]{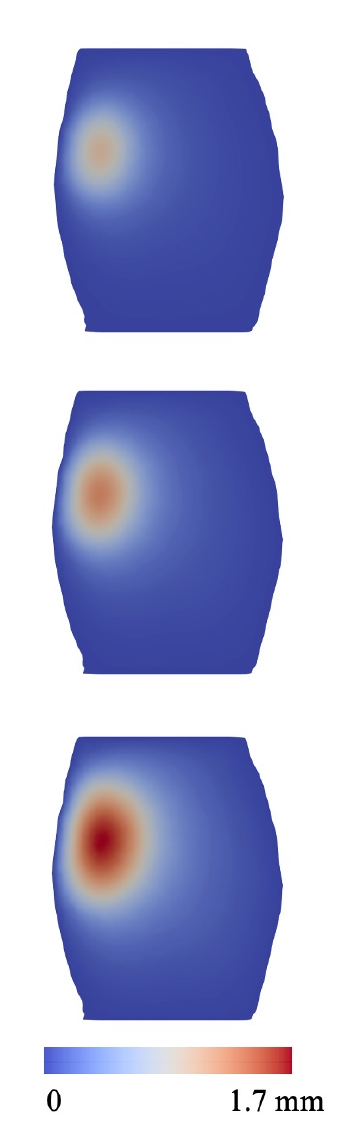}
    \caption{$\mathbf{u^s}$ for rat 4}
    \label{fig:dispdiff:d}
  \end{subfigure}
  \caption{From top to bottom: solid displacements in a coronal slice for the four rats at calibration (day 3 - first row), first (day 4 - second row) and second prediction times (day 6 - third row). Maximum displacements were observed at the second prediction time (day 6) for all rats with rat 1 exhibiting the highest displacement values.}
\label{us}
  \end{figure}

Peak displacement is observed within the tumor core for rats 1,3 and 4. In contrast, rat 2 exhibits a displacement minimum in the tumor center, which is closely encircled by regions of maximal displacement as observed in Fig.\ref{us}. Across all rats, solid displacement magnitudes increase over time. Rat 2 and 4 show moderate displacement, with maximum displacement magnitudes reaching a maximum value of $1.65$ mm for rat 4 and $1.04$ mm for rat 2.  In contrast, rats 1 and 3 show the highest displacement values, with rat 1 reaching $5.88$ mm at calibration and $9.11$ mm at the second prediction time and rat 3 reaching $5.49$ mm at the end of the simulation.

\subsubsection{Porosity}
\begin{figure}
  \centering
  \begin{subfigure}{0.27\textwidth}
    \includegraphics[width=\linewidth]{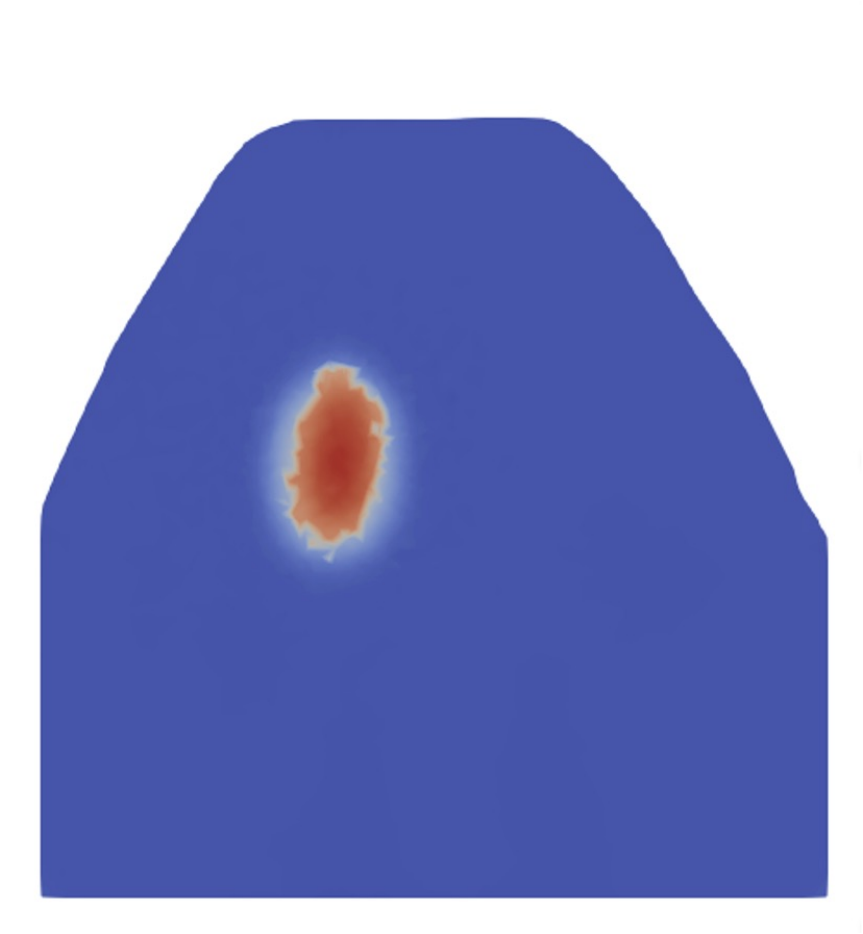}
    \caption{$\varepsilon$ at day 3}
    \label{poro_rat2_d3}
  \end{subfigure}
  \hfill
  \begin{subfigure}{0.27\textwidth}
    \includegraphics[width=\linewidth]{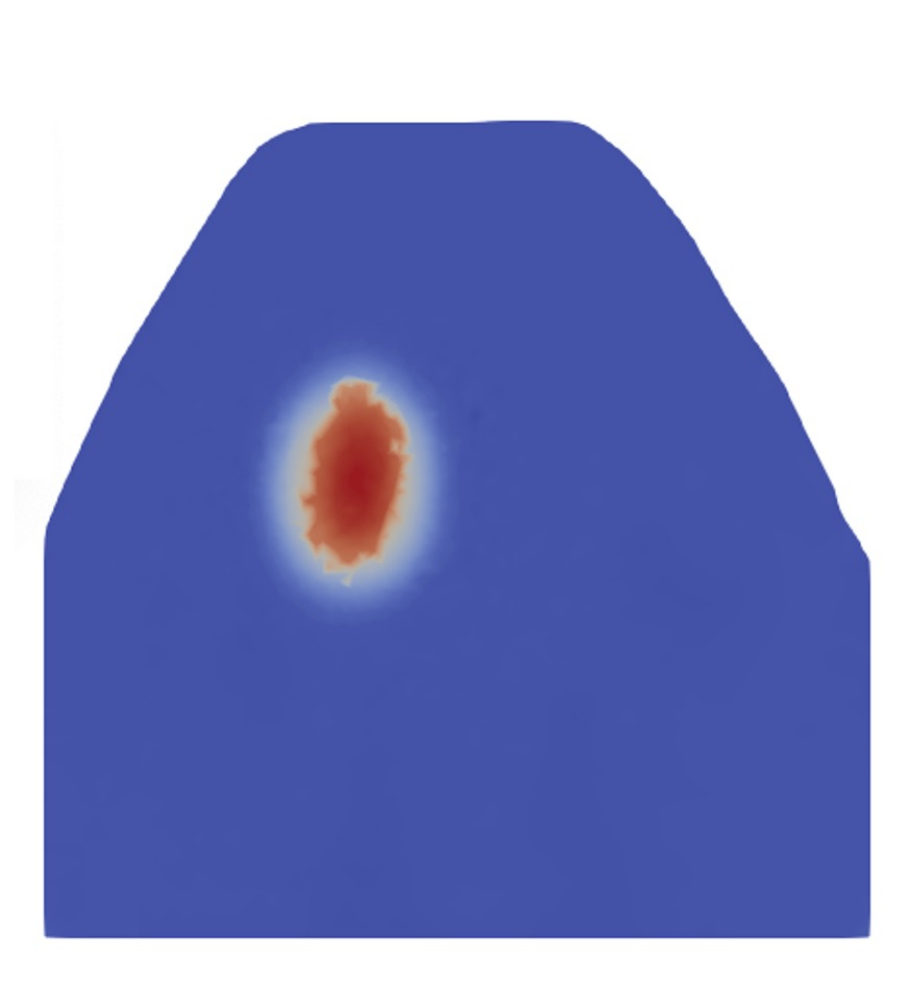}
    \caption{$\varepsilon$ at day 4}
    \label{poro_rat2_d4}
  \end{subfigure}
  \hfill
  \begin{subfigure}{0.34\textwidth}
    \includegraphics[width=\linewidth]{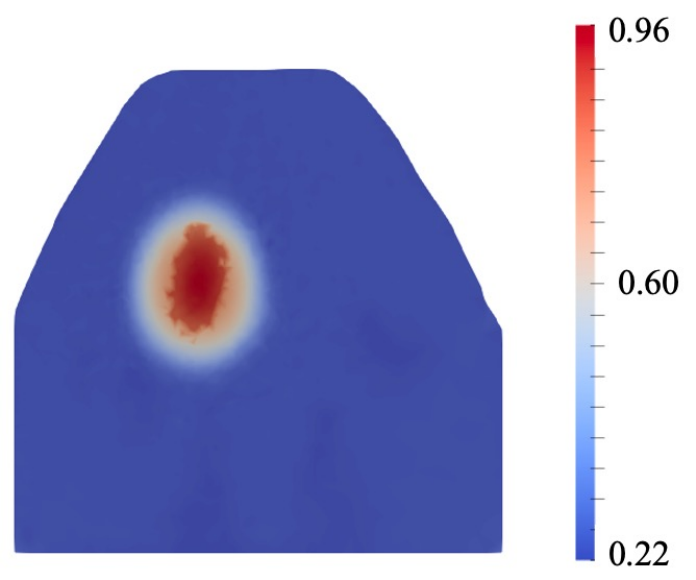}
    \caption{$\varepsilon$ at day 6}
    \label{poro_rat2_d6}
  \end{subfigure}  
  \caption{$\varepsilon$ at days 3, 4 and 6 for rat 2}
  \label{poro_rat2}
\end{figure}

For rats 2 and 4, the porosity range widened progressively from the initial interval ([0.27, 0.82]) to [0.16, 0.98] for rat 2 and [0.08, 0.95] for rat 4 at the final prediction time. Porosity evolution on one slice of rat 2 is shown in Figure~\ref{poro_rat2}. 
For rats 1 and 3, however, porosities reached non-physical values (i.e., outside the admissible interval [0,1]) during the simulation. 
To avoid this issue, in these cases the porosity was updated with the porosity results from the iteration preceding the appearance of the non-physical values,  reaching [0.025,0.99] for rat 1 and [0.002, 0.98] for rat 3.
Only less than ten element nodes reached values below 0.12 for rat 1 and 3 at the last valid porosity-update iteration. This indicates that near-zero porosity values were spatially localized. This behavior may be due to the tumor’s proximity to the boundary for rats 1 and 3, which might increase boundary-induced constraints.

\subsubsection{Tumor volume fraction}
\begin{figure}
  \centering
  \begin{subfigure}{0.26\textwidth}
    \includegraphics[width=\linewidth]{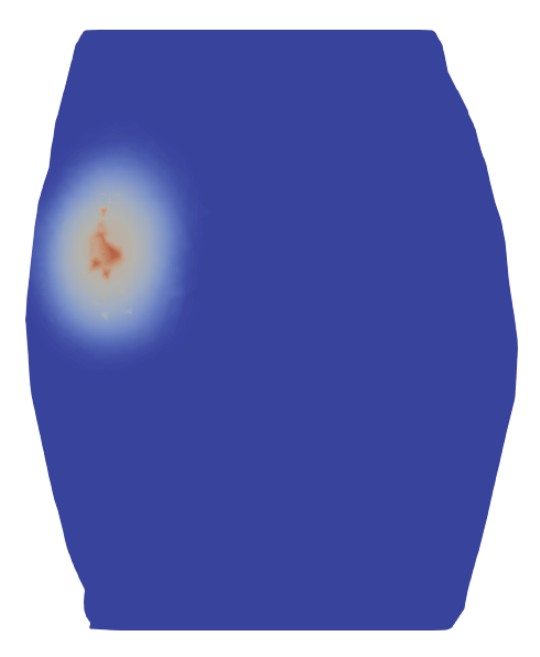}
    \caption{$\varepsilon^t$ at day 3}
    \label{epst_rat3_d3}
  \end{subfigure}
  \hfill
  \begin{subfigure}{0.26\textwidth}
    \includegraphics[width=\linewidth]{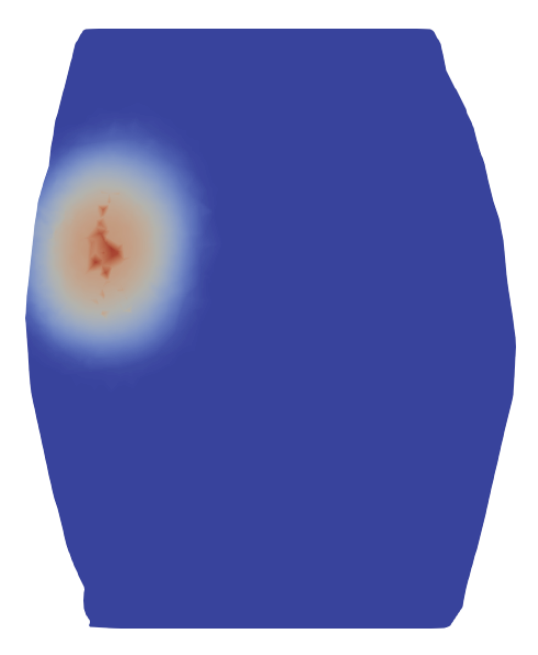}
    \caption{$\varepsilon^t$ at day 4}
    \label{epst_rat3_d4}
  \end{subfigure}
  \hfill
  \begin{subfigure}{0.35\textwidth}
    \includegraphics[width=\linewidth]{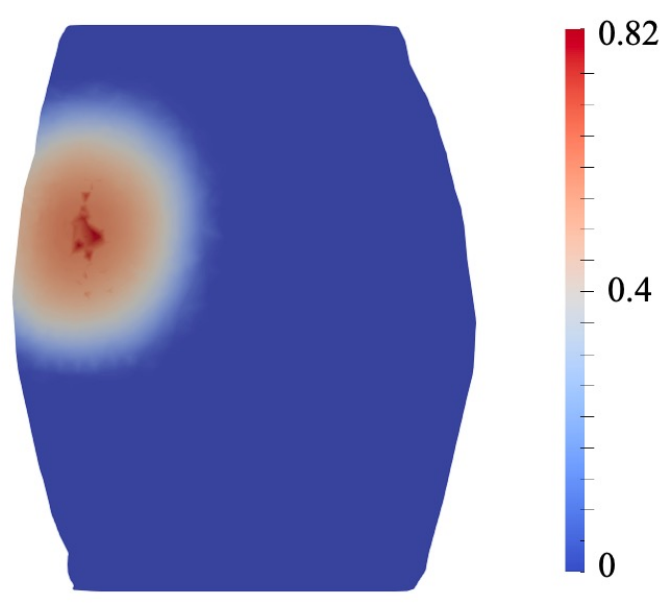}
    \caption{$\varepsilon^t$ at day 6}
    \label{epst_rat3_d6}
  \end{subfigure}  
  \caption{$\varepsilon^t$ at days 3, 4 and 6 for rat 4}
  \label{epst_rat3}
\end{figure}

The tumor volume fraction increased from calibration to the first and second prediction times. Accross all animals, the maximum tumor volume fraction started at 0.42 at the initial time and rose to values between 0.93 (rat 4) and 0.98 (rat 1) at day 6. Additionally, the values at the tumor boundaries are lower than within the core of the initial tumor, as illustrated for rat 4 in Fig.~\ref{epst_rat3}.

\subsubsection{Solid pressure}
 \begin{table}[h]
  \begin{tabular*}{\textwidth}{@{\extracolsep{\fill}}c | c c c l@{}}
    \toprule
    \makecell{rat} & \makecell{$p^s$ (KPa)\\ day 3} & \makecell{$p^s$ (KPa) \\ day 4}& \makecell{$p^s$(KPa)\\ day 6} \\
    \hline
    \hline
    \makecell{1}  & \makecell{max = 30.75\\ mean = 1.14} & \makecell{max = 40.11 \\ mean = 1.44} & 
    \makecell{max = 43.49 \\ mean = 1.78}  \\
                \hline

    \makecell{2} & \makecell{max = 12.75 \\ mean = 0.50} & \makecell{max = 15.39 \\ mean = 0.52} & \makecell{max = 17.43 \\ mean = 0.59}  \\
         \hline 
     \makecell{3} & \makecell{max = 25.79 \\ mean = 0.83} & \makecell{max = 29.17 \\ mean = 1.00} & \makecell{max = 33.05 \\ mean = 1.48} \\
    \hline 
     \makecell{4} & \makecell{max = 9.83 \\ mean = 0.53} & \makecell{max = 11.13 \\ mean = 0.57} & \makecell{max = 12.55 \\ mean = 0.65} \\
    \bottomrule
    
  \end{tabular*}
      \caption{Mean and maximum solid pressure ($p^s$) values at day 3, 4 and 6 for all rats.}
    \label{solidp}
  \end{table}

 For all rats, liquid pressure only increased by less than 1 Pa throughout the whole duration of the simulations, due to the homogeneous boundary conditions of the liquid pressure increment. As a result, the solid pressure evolution was mainly driven by the evolution of the tumor saturation and porosity values.
Hence, both mean and maximum values of solid pressure increased over time across all rats, as shown in Table~\ref{solidp} at days 3, 4, and 6. 
In particular, Rat 1 exhibited the highest solid pressure overall, with maximum values increasing from 30.75 KPa (mean 1.14 KPa) on day 3 to 40.11 KPa (mean 1.44 KPa) on day 4 and peaking at 43.49 KPa (mean 1.78 KPa) on day 6. These values indicate substantial tumor-induced stress accumulation for this rat. 
In contrast, rats 2 and 4 showed lower and more stable profiles (e.g., rat 2: maxima 12.75--17.43 KPa, means 0.50--0.59 KPa; rat 4: maxima 9.83--12.55 KPa, means 0.53--0.65 KPa), while rat 3 exhibited an intermediate increase of the solid pressure (maxima 25.79--33.05 KPa, means 0.83--1.48 KPa).

\section{Discussion}
We have introduced a poromechanical model of glioma growth which can be informed using anatomical and quantitative MRI data characterizing the architectural morphology and biological behavior of the disease. In total, twelve parameters were needed in our model and their admissible ranges were defined based on previous studies in the literature. 
Seven parameters were assigned a spatially homogeneous value across the brain domain ($c$, $\mu^t$, $\mu^l$, $\mathop \gamma\limits_{} \limits^{l \to t}$, $\nu$, $p_{start}$ and $p_{crit}$), three parameters were heterogeneously mapped from MRI data throughout the whole brain domain ($\varepsilon^t$, $\varepsilon^s\omega^{bs}$ and $E^s$, and two parameters were heterogenously defined from MRI data within the tumor region and were assigned a homogeneous value in the healthy part of the brain ($\varepsilon^l$ and $k^s_{int}$).
We identified six parameters for which data in the literature is scarce, although they have potential to have a significant contribution in driving the model dynamics (c,  $\mu^t$, $k^s_{int}$, $\mathop \gamma\limits_{} \limits^{l \to t}$, $p_{start}$ and $p_{crit}$). A variance-based sensitivity analysis helped identify two parameters that were not significantly associated with tumor growth in the rats in this study ($p_{start}$ and $p_{crit}$). 
Subsequently, only four parameters were left for animal-specific model calibration (c, $\mu^t$, $\mathop \gamma\limits_{} \limits^{l \to t}$ and $k^s_{int}$). This reduced set of parameters makes our model less dependent on longitudinal data compared to previous poroelastic formulations in the literature~\cite{sciume2013multiphase,hervas2023tumour}.

For the same experimental data, Hormuth \emph{et al.} investigated the growth of C6 glioma tumors by comparing the performance of a classical reaction-diffusion model with a mechanically-coupled reaction-diffusion models \cite{hormuth2017mechanically}. 
In their study, the model was initialized at day 10 post inoculation and parameters were calibrated using MRI data from days 12 and 14. Their parameter calibration involved one homogeneous and two voxel-wise heterogeneous parameters mapped through the spatial domain. 
The standard reaction-diffusion model generally overestimated tumor growth, yielding large and increasing relative errors in tumor volume that ranged from $16.08 \%$ up to $50.37 \%$ for a cohort of 14 rats between 15 and 20 days post inoculation (1 to 6 days after the last parameter optimization time). 
By comparison, the mechanically-coupled model substantially improved predictive accuracy, maintaining relative tumor volume errors below $8 \%$ throughout the same period. 
In our poromechanical framework, relative tumor volume errors during the prediction phase (days 4 and 6) varied from $4.73 \%$ to $36.03\%$ across the four rats considered herein, with calibration errors between $0.94\%$ and $11.27\%$. Our highest prediction error ($36.03 \%$) falls within the error range produced by the non-mechanically-coupled reaction-diffusion models in Hormuth \emph{et al.}, while our best result ($4.73 \%$) is lower than their minimum reported error.
However, the results by  Hormuth \emph{et al.} \cite{hormuth2017mechanically} are based on a more complex parametrization, which involved two voxel-wise parameters that increased the total number of parameters to at least twice the number of voxels contained in the tumor domain. Conversely, our simulations of glioma growth required a more lightweight calibration based on only four scalar parameters. This is a computational asset for striking a good balance between accuracy and efficiency, especially in modeling applications requiring multiple model calibrations (e.g., uncertainty quantification, treatment optimization).

In another work by Rey \emph{et al.}  \cite{rey2024heterogeneous}, the authors introduced an MRI-informed model for U251 orthotropic glioma using a hyperelastic constitutive model. This approach accounted for spatial variations in tissue hydraulic conductivity and porosity within the tumor by applying the \text{KC}  equation, similarly to our approach. In our study, we further informed several key model parameters and variables using clinically-available MRI measurements, namely: (i) the vascular supply volume fraction via \textit{rCBV}, which is heterogeneous across the whole tumor region of interest; (ii) the tumor volume fraction, which was initialized using \textit{ADC} maps; and (iii) different Young's modulus values in white matter and gray matter via the corresponding segmentations of these tissues on T1-weighted MRI data. However, we did not consider higher porosity in the ventricular regions, which is a feature considered in the model of Rey \emph{et al.}  that could offer greater heterogeneity in representing the poroelastic dynamics of glioma growth.

An interesting result of our study is that the variance-based sensitivity analysis showed that the starting and critical solid pressure thresholds controlling the mechanical inhibition of tumor proliferation were not associated with the predicted tumor volume. Kalli \emph{et al.}  studied the effect of proliferative and migratory ability of A172 cells (a glioblastoma cell line) and showed that the spheroids only grew signiﬁcantly until a level of around $3.5$ KPa of compressive stress \cite{kalli2019mechanical}. Additionally, previous measurements of the  tumor pressure in rat brains from the literature are in the range of 798 Pa to $3.32$ KPa \cite{boucher1997interstitial, wiig1983rat}. These values of solid pressure are overall higher than those considered for the derivation of our model (although they may be relevant in human patients  \cite{boucher1997interstitial}), thereby supporting our choice to ignore mechanical inhibition of proliferation in our model applied to C6 glioma growth.

Previous studies have shown measurements of the tumor intracranial fluid pressure (TIFP), intracranial pressure (ICP), and solid stress that are on average similar to those obtained in our study for solid pressure.
For example, Boucher \emph{et al.} \cite{boucher1997interstitial} measured ICP values in and outside F98 and R3230AC tumors in Fisher rat models and obtained an average of $1.89$ KPa for F98 and $2.96$ KPa (both parenchymal and pial surface tumors). 
Elmghirbi \emph{et al.} \cite{elmghirbi2018toward} measured the TIFP in 36 athymic female rats with U251 tumors using the wick-in-needle technique and obtained a sample mean TIFP of $0.79$ KPa. Moreover, Rey \emph{et al.} \cite{rey2024heterogeneous} numerically obtained a stress in the solid phase up to $0.5$ KPa. Compared to all these lower values from the literature, we obtained a maximum solid pressure that is three orders of magnitude higher. Nevertheless, the comparison of our ﬁndings with those of Rey \emph{et al.} requires careful consideration due to the different assumptions inherent in both modeling approaches. In our study, the solid pressure is deﬁned as a weighted sum of both the interstitial liquid pressure and the tumor phase pressure. In contrast, Rey \emph{et al.} only include the interstitial liquid in the porosity, treating the tumor-induced pressure as distinct from the fluid phase pressure. Furthermore, their investigation focuses on a parametric analysis of solid phase stress driven by high interstitial fluid pressure and tumor growth, particularly concerning the effects of leaky vasculature and stiffness levels. Therefore, the primary focus of their study was not on real-time predictions. Additionally, the Young’s moduli that they considered in the healthy and tumor tissue is considerably higher than our and other studies \cite{hormuth2017mechanically,christ2010mechanical}, as they employed values within the range [$2.75$ ; $12.96$] KPa and [$1.56$ ; $15.90$] KPa, respectively. All these examples highlight that making a direct comparison of solid pressure levels between models is complicated and heavily relies on the intent of the study as well as model definition and parameterization.

Despite the promising results obtained in this preliminary study, we also note some limitations. First, the peak pressures obtained in our study point to a potential limitation in the formulation of the pressure saturation law. Since the solid pressure is linked to saturation through a phenomenological relationship using a tangent function (see Eq.~\eqref{ps}), high saturation values may produce unrealistically large peak pressures. This numerical behavior arises because the tangent function exhibits rapid growth as its argument approaches $\frac{\pi}{2}$, amplifying pressures at near-complete saturation. To address this issue, experimental tests would need to be performed on the speciﬁc cell line to further refine the expression as advised in \cite{sciume2013multiphase}. 
Second, we assumed that the direction of tumor growth would be heavily influenced by tissue vascularization, which is represented through parameter $\varepsilon^s\omega^{bs}$ in the mass transfer function and is informed using rCBV measurements. However, the results of our model simulations and the MRI measurements of the tumor region suggest that this is not always the case (see Fig.~\ref{ratwbs}). Future studies should investigate whether and how other parameters might be responsible for the tumor growth directions, such as permeability  and tumor dynamic viscosity.
Third, permeability was defined as a heterogeneous spatial map only within the tumor, using the KC equation and the $v_e$ maps. However, the link between permeability and the $v_e$ or \textit{ADC} maps needs to be further investigated outside the tumor, although results in the literature are contradictory. While Mui \emph{et al.} \cite{mui2022correlations} found a significant positive correlation between $v_e$ and \textit{ADC} maps in a study in 20 patients with nasopharyngeal carcinoma, Mills \emph{et al.} \cite{mills2010candidate} found no correlation between the $v_e$ and \textit{ADC} maps in a voxel-by-voxel analysis or comparison of median values in 19 patients with glioblastoma. These latter results suggest that the \textit{ADC} and $v_e$ maps should both be explored to inform different parameters of tumor growth as they reflect different but complementary aspects of the tumor microenvironment. In fact, Vajapeyam \emph{et al.} showed that combined together, \text{ADC} and permeability maps can help differentiate high-grade from low-grade pediatric brain tumors \cite{vajapeyam2018multiparametric}, highlighting the added diagnostic value of multiparametric imaging. Finally, more MRI data types that are available in clinical scenarios could be investigated to better inform the model. For instance, flAIR combined with T2-weighted MRI can better differentiate between edema and cerebrospinal fluid, while T1GD can detect blood brain barrier (BBB) disruption and provide information on leaky vasculature. Moreover, ADC values outside the tumor can inform on the volume fraction of healthy brain cells.

Therefore, the promising prediction results of this preliminary study and the reduced set of sensitive parameters that require calibration with longitudinal data suggest that our proposed model could be a promising candidate to investigate personalized tumor forecasting in glioma patients. This could constitute an important advance in brain cancer forecasting, since existing models either rely on temporally-resolved ODE formulations \cite{plaszczynski2023predicting,bruningk2021intermittent,delobel2023overcoming, iarosz2015mathematical} or reaction-diffusion models \cite{wang2009prognostic, jackson2015patient, hormuth2017mechanically, balcerak2025individualizing, lipkova2019personalized, harkos2022inducing} that offer limited insight into the complex interplay between tumor dynamics, tissue architecture (e.g., porosity, relative solid and fluid composition), and mechanical deformations. The capability of predicting these features for  host and tumor tissue could potentially enable more insightful analysis of biological mechanisms underlying growth and treatment response in experimental studies \cite{gevertz2024minimally, segura2022optimal,banga2025mechanistic} and for individual patients \cite{yankeelov2013clinically,lorenzo2025validating,corwin2013toward}. The latter could therefore offer the possibility to design better therapeutic strategies to combat disease progression, and extend survival to the disease \cite{chaudhuri2023predictive,lipkova2019personalized,colombo2015towards}.

\section{Conclusion}

We propose a three-phase poromechanical model to predict glioma growth, whose dynamics are primarily controlled by four scalar parameters (c, $\mu^t$, $\mathop \gamma\limits_{} \limits^{l \to t}$ and $\alpha^k$). We demonstrate that the main variables of the model and the four driving parameters can be informed by standard, longitudinal anatomical and quantitative MRI data. Although we obtained a promising predictive performance in a preliminary tumor forecasting study in a small cohort of n=4 rats, further model development, experimental work, and validation are needed to better characterize poromechanical parameters and improve how diverse MRI data types can inform the model components. Nevertheless, this study constitutes a first step towards the development of a patient-specific model accounting for the multiscale, spatiotemporal poromechanics underlying the dynamics of glioma progression and therapeutic response.

\section{Acknowledgments}
The authors thank the Oden institute for providing the MRI data. MAA was supported by the Institute for Advanced Studies (IAS) of the University of Luxembourg, project U-AGR-6046-00-B. SU thanks the Agence National de la Recherche (ANR) (France) and the Fond National de la Recherche (FNR) (Luxembourg) joint grant number ANR-21-CE45-0025-01. DAH thanks the Cancer Prevention and Research Institute of Texas for support through CPRIT RP220225, the National Science Foundation through DMS 2436599, and the American Cancer Society via IRG-21-135-01-IRG. 
GL acknowledges the support of a fellowship from ‘‘la Caixa” Foundation (ID 100010434). The fellowship code is LCF/BQ/PI23/11970033.
GL also acknowledges grant PID2023-146347OA-I00 funded by $MICIU/AEI/10.13039/501100011033$ and ERDF/EU, as well as grant $RYC2022-036010-I$ funded by $MICIU/AEI/10.13039 \\/501100011033$ and $ESF+$.
TEY thanks the US National Cancer Institute for funding through $1R01CA260003$.  The results presented in this article were carried out using the HPC facilities of the University of Luxembourg~\cite{VBCG_HPCS14} (see \url{https://hpc.uni.lu}). The authors also thank Dr. Camilo Afanador, Dr. Thomas Lavigne and Dr. Jack Hale for their valuable help throughout the course of this work.
 \appendix
 \section{Variational formulation of the coupled system}\label{Appendixvar}

 This appendix introduces the weak formulation of Eqs.~\eqref{4},~\eqref{5},~\eqref{6} and~\eqref{momentum} for the application of the FE method. Of note Eq.~\eqref{4} is implicitly solved in the weak formulations of Eqs.~\eqref{5} and~\eqref{6}. Considering $q^l$, $q^t$, and $\mathbf{v}$ to be the test functions; $S^t_n$, $p^l_n$, $\mathbf{u^s_n}$ and $\varepsilon_n$, the solutions of the previous time step; and $dS^t$, $dp^l$, and $\mathbf{du^s}$ the increments of the variables, the variational formulation of Eq.~\eqref{5} becomes:
 \begin{equation}
 \begin{aligned}
 \int_\Omega \frac{1}{\Delta t} \left( (S^t_n+dS^t) \nabla \cdot \mathbf{du}^s\right) q_t \, d\Omega 
 +\int_\Omega \frac{1}{\Delta t} \left( \varepsilon dS^t \right) q_t \, d\Omega 
 \\
 +\int_\Omega \frac{1}{\Delta t} \left( \alpha_k \frac{k^s_{int}}{\mu^t}\right)  \nabla\left(p^l_n + dp^l+c\tan\left(\frac{\pi}{2} (S^t_n+dS^t) \right)\right) \cdot \nabla q_t\, d\Omega 
 \\
 -\int_\Omega \mathop {\mathop \gamma\limits_{} }\limits^{l \to t}\, w^{bs} \, \varepsilon \, \big( S^t_n + dS^t \big) q_t \, d\Omega = 0
 \end{aligned}
     \label{var5}
 \end{equation}

and Eq.~\eqref{6} becomes: 
\begin{equation}
\begin{aligned}
\int_\Omega \frac{1}{\Delta t} \left( (1 - (S^t_n+dS^t) \nabla \cdot \mathbf{du}^s  \right) q_l \, d\Omega 
-\int_\Omega \frac{1}{\Delta t} \left( \varepsilon dS^t \right) q_l \, d\Omega 
\\
+\int_\Omega \frac{1}{\Delta t} \left( \alpha_k \frac{k^s_{int}}{\mu^l}\right)  \nabla\left(p^l_n + dp^l \right) \cdot \nabla q_l\, d\Omega 
+\int_\Omega \mathop {\mathop \gamma\limits_{} }\limits^{l \to t}\, w^{bs} \, \varepsilon \, \big( S^t_n + dS^t \big) q_l \, d\Omega = 0.
\end{aligned}
    \label{var6}
\end{equation}
Finally, Eq.~\eqref{momentum} becomes:
\begin{equation}
\begin{aligned}
& \frac{1}{\Delta t} \bigg( 2 \mu \, \operatorname{sym}\big(\nabla (\mathbf{u}^s_n + d\mathbf{ u^s})\big) : \operatorname{sym}\big(\nabla \mathbf{v}\big) \\
& \quad + \lambda_s \nabla \cdot (\mathbf{u}_n + d\mathbf{ u}) \, \nabla \cdot \mathbf{v} - p^s(p^l_n + dp^l,\ S^t_n + d S^t) \nabla \cdot \mathbf{v} \bigg) d\Omega = 0,
\end{aligned}
\end{equation}
where $\operatorname{sym}$ is the symmetric part of the displacement gradient: 
\begin{equation}
\varepsilon(\mathbf{u}) = \operatorname{sym}(\nabla \mathbf{u}) = \frac{1}{2}\left( \nabla \mathbf{u} + (\nabla \mathbf{u})^T \right).
\end{equation}

\bibliographystyle{unsrt}

\bibliography{BMbib}

\end{document}